\documentclass[11pt]{article}
\usepackage[numbers]{natbib}
\usepackage{graphics}
\usepackage{color}
\usepackage{bbm}
\usepackage{amsmath}
\usepackage{hyperref}

\definecolor{fgred}{rgb}{0.8,0,0}     
\definecolor{fgblue}{rgb}{0,0,1}      

\newcommand{\modified}[1]{{#1}}

\newcommand{\bfm}[1]{\textbf{\em #1}}
\newcommand{\bfs}[1]{\boldsymbol{#1}}
\newcommand{\brm}[1]{{\bf #1}}

\graphicspath{{../Fig/}{../Fig_Implicit/}{../Fig_Xfig/}{../Fig_AP/}}

\newcommand{\inclps}[3]{\resizebox{#1}{#2}{\includegraphics{#3}}}

\makeatletter
\def\blfootnote{\xdef\@thefnmark{}\@footnotetext}
\makeatother

\begin{document}

\title{Implicit yield function formulation for granular \\ and rock-like materials}

\author{S. Stupkiewicz, R. Denzer, A. Piccolroaz, D. Bigoni}


\date{}

\maketitle

\begin{abstract}
The constitutive modelling of granular, po\-rous and quasi-brittle materials is
based on yield (or damage) functions, which may exhibit features (for instance, lack
of convexity, or branches where the values go to infinity, or \lq false elastic
domains') preventing the use of efficient return-mapping integration schemes. This
problem is solved by proposing a general construction strategy to define an
implicitly defined convex yield function starting from any convex yield surface.
Based on this implicit definition of the yield function, a return-mapping
integration scheme is implemented and tested for elastic-plastic (or -damaging) rate
equations. The scheme is general and, although it introduces a numerical cost when
compared to situations where the scheme is not needed, is demonstrated to perform
correctly and accurately.

\vspace{1ex}
\end{abstract}

\blfootnote{
\\[-3mm]
S. Stupkiewicz \\
Institute of Fundamental Technological Research (IPPT), \\
Polish Academy of Sciences, Pawinskiego 5b, 02-106 Warsaw, Poland \\
E-mail: sstupkie@ippt.pan.pl \\[1ex]
R. Denzer \\
TU Dortmund, Leonhard-Euler-Strasse 5, D-44227 Dortmund, Germany \\
E-mail: ralf.denzer@tu-dortmund.de \\[1ex]
A. Piccolroaz, D. Bigoni \\
University of Trento, via Mesiano 77, I-38123 Trento, Italy \\
E-mail: roaz@ing.unitn.it, bigoni@unitn.it
}

\section{Introduction}

The pressure-sensitive yielding of granular, porous, and quasi-brittle materials
(such as ceramic or metallic powders, porous metals, rocks and concretes), or the
Lode's angle dependence of high-strength and shape memory alloys, forces the use of
complex yield functions in the plastic or damaging constitutive  modelling. These
yield functions (three examples are those introduced by Jere\-mic et
al.~\cite{Jeremic99}, by Foster et al.~\cite{Foster05} and by Bigoni and
Piccolroaz~\cite{BigoniPiccolroaz04}, the last called \lq BP' in the following)
often display \lq undesidered features', such as for instance lack of convexity, or
regions where they blow up to infinity or, as indicated by Brannon and
Leelavanichkul~\cite{BrannonLeelavanichkul10}, \lq false elastic domains' (in which
negative values are associated to stress states external to the \lq true' elastic
domain), preventing the use of standard return-mapping integration algorithms.
Moreover, these yield functions often describe high-curvature surfaces and nearly
sharp vertices, which are difficult to treat with the necessary accuracy and
definitely slow down numerical procedures. Remedies to these problems have been
proposed in \cite{BrannonLeelavanichkul10} and \cite{Penasa14}, but the former
technique, based on a multi-stage algorithm, is very complex, while the latter,
based on a cutoff-substepping return algorithm, is applicable only to the BP yield
function.

The purpose of the present article is to introduce a new approach to the problem,
where a general procedure to construct an implicit definition of a convex yield
function starting from a convex yield surface is proposed and the related
application within a standard return-mapping scheme is explained and tested. The
method is general, simple to be implemented, and applies to every convex yield
surface. It is shown to be, on the one hand, associated to a computational cost
(which has to be regarded as the counterpart of the complexity of the employed yield
function), but on the other hand, to be robust and to provide accurate and stable
results.


The paper is organized as follows. The general setting of the implicit yield
function formulation is presented in Section~\ref{sec:general}. The original BP
yield function \cite{BigoniPiccolroaz04} is recalled, and its undesired features are
illustrated in Section~\ref{sec:BP}. In Section~\ref{sec:implicit}, the general
formulation is specified for the BP yield surface: the implicit yield function is
formulated in the $(p,q)$-space, and the presentation style is oriented towards
computer implementation employing an automatic differentiation (AD) technique and
\emph{AceGen}, an automatic code generation system \cite{Korelc02,Korelc09}. The
complete \emph{AceGen} input is also provided as a supplementary material
accompanying the paper along with ready-to-use subroutines in \emph{C},
\emph{Fortran}, \emph{Mathematica} and \emph{Matlab}. Feasibility of the implicit
yield function concept in incremental elastoplasticity is finally illustrated in
Section~\ref{sec:examples}.

\section{Implicit yield function: general setting}
\label{sec:general}

Consider a yield function $F(\bfs{\sigma},\bfs{\eta})$, depending on stress
$\bfs{\sigma}$ and hardening variables $\bfs{\eta}$, that defines a convex elastic
domain ${\cal E}$ such that the boundary $\partial{\cal E}$ of the elastic domain,
i.e., the yield surface, is specified by the zero level set,
$F(\bfs{\sigma},\bfs{\eta})=0$. As discussed in the introduction, the yield function
$F$ may be non-convex or defined infinity in some regions, see Section~\ref{sec:BP}
for the illustration of those features in the case of the BP yield function.


The strategy proposed in the present paper to overcome the difficulties mentioned
above is to introduce an alternative yield function
$F^\ast(\bfs{\sigma},\bfs{\eta})$ that is defined for an arbitrary stress, and such
that its zero level set $F^\ast=0$ (i.e., the yield surface) is identical to that of
the original yield function, $F=0$. In this way, the elastic domain ${\cal E}$ and
its closure, which are the only two relevant quantities from the mechanical point of
view, will remain the same as for the original yield function
$F(\bfs{\sigma},\bfs{\eta})$, but convexity and finiteness will be enforced.

\modified{Note that the yield surface may be non-convex with respect to the
hardening or damage variables $\bfs{\eta}$ which could result in convergence
problems in an incremental computational scheme, e.g., \cite{ShengAugardeAbbo11}.
However, such non-convexity would be a constitutive feature of a specific
hardening/softening law that cannot be removed by reformulating the yield function
itself.} For the sake of a compact notation, we will skip the dependence of $F$ on
$\bfs{\eta}$ in this section, because only the dependency on $\bfs{\sigma}$ is
relevant in the following.

\begin{figure}
  \centerline{\scalebox{0.9}{\input{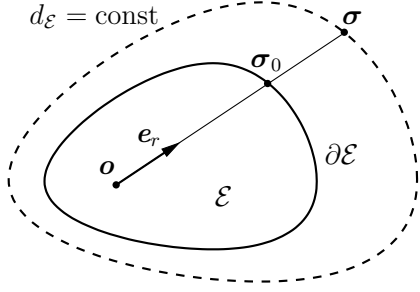}}}
  \vspace{1ex}
  \caption{Convex distance function.
           \label{fig:distance}}
\end{figure}

The main idea is to use the so-called convex distance function $d_{\cal
E}(\bfs{\sigma})$ of the zero level set $F=0$, i.e., the boundary of the elastic
domain $\partial{\cal E}$, of the original yield function. The convex distance
function is defined as
\begin{equation}
d_{\cal E}(\bfs{\sigma}) :=
  \frac{\|\bfs{\sigma} - \bfm{o}\|}{\|\bfs{\sigma}_0(\bfs{\sigma}) - \bfm{o}\|}
\end{equation}
with $\|\cdot\|$ being the 2-norm, $\bfm{o}\in{\cal E}\setminus\partial{\cal E}$
representing a fixed reference point inside the elastic domain, and we define
$d_{\cal E}(\bfm{o}) = 0$. The convex distance function has the property
\begin{equation}
d_{\cal E}(\bfs{\sigma}) = \begin{cases}
                [0, 1)          & \text{if\quad} \bfs{\sigma} \in {\cal E} \setminus \partial{\cal E} \\
                1               & \text{if\quad} \bfs{\sigma} \in \partial{\cal E} \\
                > 1             & \text{if\quad} \bfs{\sigma} \not\in {\cal E}
               \end{cases}
\end{equation}
and it is a convex function, see \cite{BonnesenFenchel87,HiriartUrrutyLemarechal93}
for a proof. In this definition, $\bfs{\sigma}_0(\bfs{\sigma})$ is the unique
intersection of a ray from $\bfm{o}$ in the direction of $\bfs{\sigma}$ with the
boundary $\partial{\cal E}$ of the elastic domain as depicted in Fig.\
\ref{fig:distance}.

To compute the unique intersection point $\bfs{\sigma}_0(\bfs{\sigma})$ we define a
ray from the reference point $\bfm{o}$ in the direction of $\bfs{\sigma}$ by
\begin{equation}
  \bfm{r}(\bfs{\sigma}, \hat{\varrho})
    = \bfm{o} + \hat{\varrho} \bfm{e}_r
\end{equation}
with
\begin{equation}
  \bfm{e}_r = \frac{\bfs{\sigma} - \bfm{o}}{\|\bfs{\sigma} - \bfm{o}\|} ,
    \quad \hat{\varrho} > 0 \quad {\rm and} \quad \bfs{\sigma} \neq \bfm{o} ,
\end{equation}
and introduce the following abbreviations
\begin{equation}
  \bfs{\varrho} = \bfs{\sigma}-\bfm{o} , \quad
  \bfs{\varrho}_0 = \bfs{\sigma}_0-\bfm{o} , \quad
  \varrho = \|\bfs{\varrho}\| , \quad
  \varrho_0=\|\bfs{\varrho}_0\| ,
\end{equation}
so that $\bfm{r}(\bfs{\sigma},\varrho)=\bfs{\sigma}$ and
$\bfm{r}(\bfs{\sigma},\varrho_0)=\bfs{\sigma}_0$.

The intersection of the ray with the boundary $\partial{\cal E}$ of the elastic
domain, i.e.\ the original yield surface, is computed by
\begin{equation}
 \label{eq:lambda}
 F_0(\varrho_0,\bfs{\sigma}) = F(\bfm{r}(\bfs{\sigma}, \varrho_0)) = 0 ,
\end{equation}
which, for a fixed $\bfs{\sigma}$, is in general a nonlinear equation for
$\varrho_0$. One would solve this nonlinear equation, e.g., by a Newton-type or a
regula falsi method. If $\varrho_0$ is the unique solution of Eq.\ \ref{eq:lambda},
then the intersection point is given by
\begin{equation}
 \label{eq:XI}
 \bfs{\sigma}_0(\bfs{\sigma}) = \bfm{r}(\bfs{\sigma}, \varrho_0(\bfs{\sigma})) ,
\end{equation}
where we indicate that the solution $\varrho_0$ depends (implicitly) on
$\bfs{\sigma}$ through Eq.\ (\ref{eq:lambda}).

With this at hand, we can implicitly define a new convex yield function $F^*$ based
only on the yield surface $F = 0$ of the original yield function by
\begin{equation}
 \label{eq:Fs:gen}
 F^\ast(\bfs{\sigma}) =
  d_{\cal E}(\bfs{\sigma}) - 1  =
  \frac{\varrho(\bfs{\sigma})}{\varrho_0(\bfs{\sigma})} - 1 ,
\end{equation}
where the term $-1$ is introduced such that we obtain the common property $F^\ast=0$
on the yield surface. One should observe that the zero level sets $F^\ast=0$ and
$F=0$ are identical but that the new yield function $F^\ast$ inherits the convexity
from the convex distance function $d_{\cal E}$.

In plasticity, we usually need the first derivative of the yield function with
respect the the stress $\bfs{\sigma}$, e.g., to define the flow direction in the
case of an associated flow rule, and with respect to the hardening variables
$\bfs{\eta}$, and often also the second derivative, e.g., to compute the consistent
tangent. By differentiating Eq.~(\ref{eq:Fs:gen})$_2$, we have
\begin{equation}
  \label{eq:dFs1}
  \frac{\partial F^\ast}{\partial\bfs{\sigma}} =
    \frac{1}{\varrho_0} \frac{\partial \varrho}{\partial\bfs{\sigma}} -
    \frac{\varrho}{\varrho_0^2} \frac{\partial \varrho_0}{\partial\bfs{\sigma}} ,
\end{equation}
where the explicit derivative $\partial\varrho/\partial\bfs{\sigma}$ reads
\begin{equation}
  \label{eq:drho}
  \frac{\partial \varrho}{\partial\bfs{\sigma}} = \bfm{e}_r .
\end{equation}
The dependence of $\varrho_0$ on $\bfs{\sigma}$ is implicit through
Eq.~(\ref{eq:lambda}). To compute the derivative of this implicit dependence, the
total derivative of Eq.~(\ref{eq:lambda}) with respect to $\bfs{\sigma}$ is
computed,
\begin{equation}
\label{eq:dF0:gen}
  \frac{\partial}{\partial\bfs{\sigma}} F(\bfm{o} + \varrho_0(\bfs{\sigma})
    \bfm{e}_r(\bfs{\sigma})) =
        \left( \varrho_0 \frac{\partial\bfm{e}_r}{\partial\bfs{\sigma}} +
    \frac{\partial\varrho_0}{\partial\bfs{\sigma}} \otimes \bfm{e}_r \right)
    \left[ \left. \frac{\partial F}{\partial\bfs{\sigma}} \right|_{\bfs{\sigma}_0} \right]
    = 0 ,
\end{equation}
where the gradient of the original yield function $\partial F/\partial\bfs{\sigma}$
is evaluated at $\bfs{\sigma}_0$, and we have
\begin{equation}
  \label{eq:der}
  \frac{\partial\bfm{e}_r}{\partial\bfs{\sigma}} = \frac{1}{\varrho}
    ( \mathbbm{S} - \bfm{e}_r \otimes \bfm{e}_r )
\end{equation}
with $\mathbbm{S}$ denoting the symmetrizing fourth order tensor. In
Eq.~(\ref{eq:dF0:gen}) and in the following, the square brackets indicate double
contraction of the second order tensor enclosed within the brackets with the fourth
order tensor preceding the brackets. Equation (\ref{eq:dF0:gen}) is solved for the
implicit derivative $\partial\varrho_0/\partial\bfs{\sigma}$, thus yielding
\begin{equation}
  \label{eq:drho0}
  \frac{\partial\varrho_0}{\partial\bfs{\sigma}} = \frac{\varrho_0}{\varrho}
    \bfm{e}_r - \frac{\varrho_0}{\varrho} \left( \bfm{e}_r \cdot
    \left. \frac{\partial F}{\partial\bfs{\sigma}} \right|_{\bfs{\sigma}_0} \right)^{-1}
    \left. \frac{\partial F}{\partial\bfs{\sigma}} \right|_{\bfs{\sigma}_0} .
\end{equation}
Finally, combing Eqs.~(\ref{eq:dFs1}), (\ref{eq:drho}) and (\ref{eq:drho0}), the
first derivative of $F^\ast$ with respect to $\bfs{\sigma}$ is obtained as
\begin{equation}
  \label{eq:dFs}
  \frac{\partial F^\ast}{\partial\bfs{\sigma}}
   = \frac{1}{\varrho_0}
      \left( \bfm{e}_r \cdot
      \left. \frac{\partial F}{\partial\bfs{\sigma}} \right|_{\bfs{\sigma}_0}
      \right)^{-1}
      \left. \frac{\partial F}{\partial\bfs{\sigma}} \right|_{\bfs{\sigma}_0} .
\end{equation}
The main ingredient of $\partial F^\ast/\partial\bfs{\sigma}$ is the gradient of the
original yield function $\partial F/\partial\bfs{\sigma}$ evaluated at the
intersection point $\bfs{\sigma}_0$, and it is seen that the former is equal to the
latter multiplied by a scalar.

For completeness, the second derivative of $F^\ast$ with respect to $\bfs{\sigma}$
is provided in Appendix~\ref{app:second}. The derivatives with respect to the
hardening variables $\bfs{\eta}$ and the mixed second derivatives can be obtained
analogously.
In the implementation approach adopted in the present work, the necessary
derivatives of the implicit yield function $F^\ast$ are obtained using an automatic
differentiation technique, hence the explicit formulae are, in fact, not needed. The
corresponding formulation, with application to the BP yield function, is presented
in Section~\ref{sec:implicit}.


\section{Original BP yield function}
\label{sec:BP}


In this section, the original Bigoni--Piccolroaz (BP) yield function
\cite{BigoniPiccolroaz04} is briefly introduced, and its deficiencies concerning its
practical application in computational plasticity are illustrated.

Introduce first the following invariants of the stress tensor $\bfs{\sigma}$:
  \begin{equation}
    p = -\frac{1}{3} \, \mbox{tr}\bfs{\sigma} , \quad
    q = \sqrt{3 J_2} , \quad
    \theta = \frac{1}{3} \, \cos^{-1} \left( \frac{3\sqrt{3}}{2} \,
      \frac{J_3}{J_2^{3/2}} \right) ,
  \end{equation}
where $\theta$ is the Lode angle, and
$J_2=\frac{1}{2}\bfs{\sigma}'\cdot\bfs{\sigma}'$ and $J_3=\det\bfs{\sigma}'$ are the
usual invariants of the stress deviator
$\bfs{\sigma}'=\bfs{\sigma}+\frac{1}{3}p\bfm{I}$, see, for instance,
\cite{Bigoni12}.

The original
BP yield function is defined by the following equations:
  \begin{equation}
    \label{eq:F}
    F(\bfs{\sigma},\bfs{\eta}) = \hat{F}(p,q,\theta,\bfs{\eta}) = f(p) + \frac{q}{g(\theta)} ,
  \end{equation}
where
  \begin{equation}
    \label{eq:f}
    f(p) = \left\{
      \begin{array}{ll}
        -M p_c \sqrt{(\Phi-\Phi^m)[2(1-\alpha)\Phi+\alpha]} \;\; & \mbox{if} \; \Phi \in [0,1], \\
        +\infty & \mbox{otherwise},
      \end{array} \right.
  \end{equation}
  \begin{equation}
    \Phi = \frac{p+c}{p_c+c} ,
  \end{equation}
and
  \begin{equation}
    g(\theta) = \frac{1}{\cos[\beta\pi/6 - (1/3) \cos^{-1}(\gamma \cos 3\theta)]} .
  \end{equation}
Here, $p_c$ and $c$ define the size and position of the yield surface along the
hydrostatic axis, and $M$, $m$, $\alpha$, $\beta$ and $\gamma$ are parameters that
define the shape of the yield surface in the stress space. All these parameters may
depend on the hardening variables that are here left unspecified and collectively
denoted by $\bfs{\eta}$. For instance, in the model for ceramic powder compaction
\cite{Piccolroaz06-1,Piccolroaz06-2}, the forming pressure $p_c$ is adopted as the
only hardening variable that is related to the volumetric plastic deformation,
cohesion $c$ is assumed to depend on $p_c$, while the remaining parameters are
assumed constant.

\begin{figure}[!htcb]
  \centerline{
    \begin{tabular}{l}
      (a) 
      \inclps{!}{0.38\textwidth}{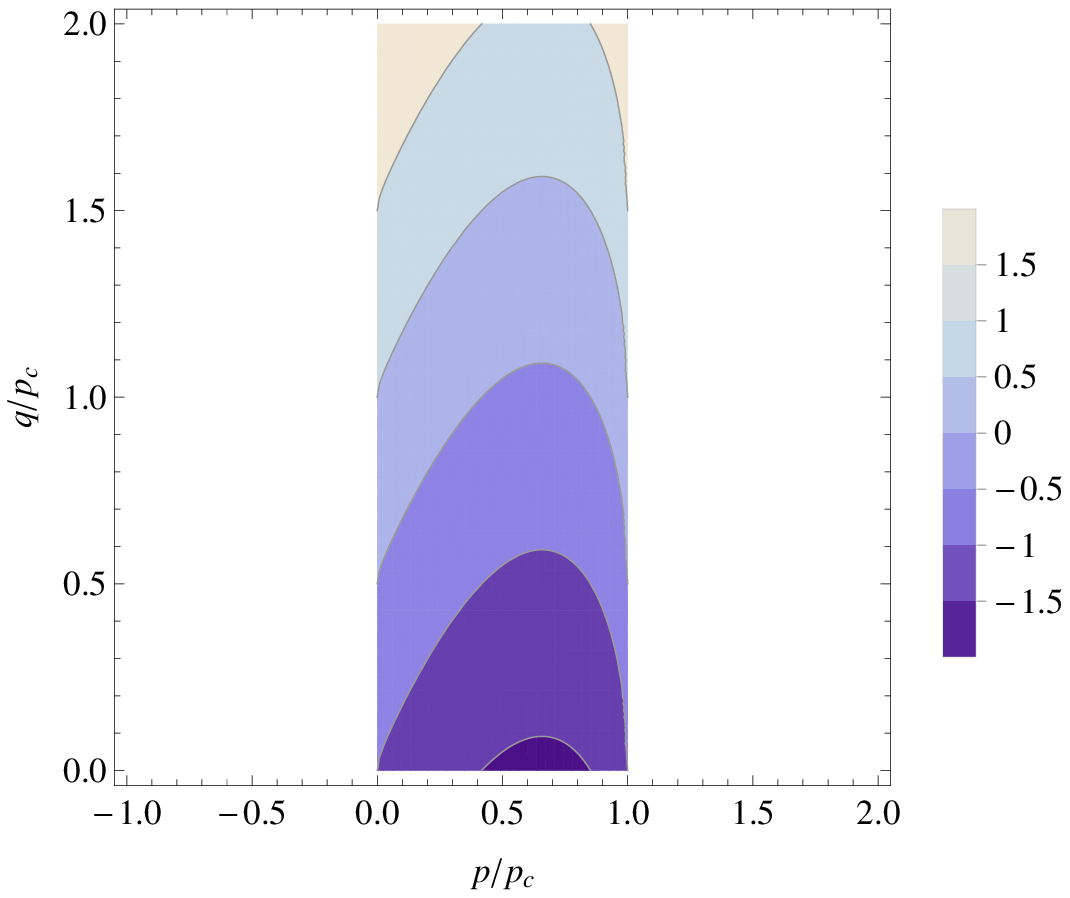} 
      (b) 
      \inclps{!}{0.38\textwidth}{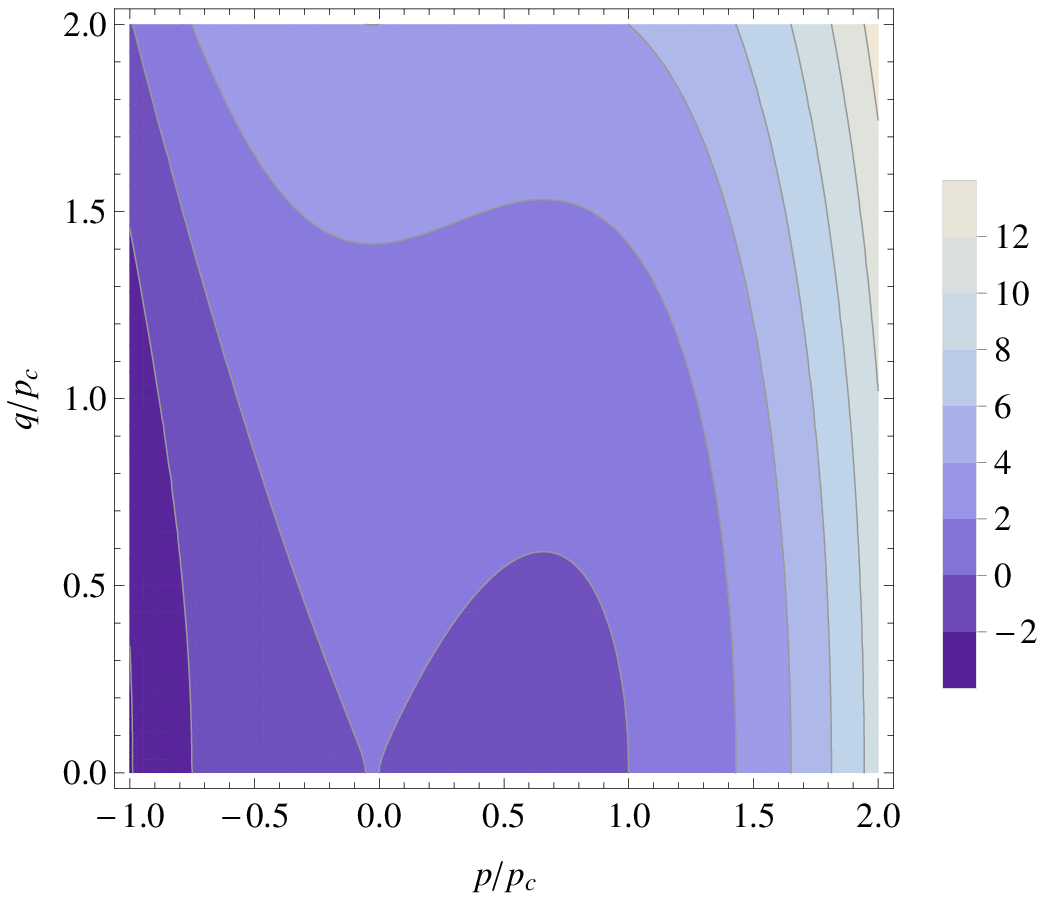}
    \end{tabular}
    }
  \caption{Iso-lines of the BP yield function in the $(p,q)$--space: (a) original
           yield function $F$ is defined infinity in the white regions,
           (b) the squared yield function $F_2$, Eq.~(\ref{eq:squared}), is not convex
           ($F$ is normalized by $p_c$ and $F_2$ is normalized by $p_c^2$).
           Parameters of the yield function correspond to alumina powder, see
           Section~\ref{sec:examples}, and Lode angle is assumed as $\theta=0$.
           \label{fig:yield}}
\end{figure}

To enforce convexity, the BP yield function in its original version (\ref{eq:F}) is
defined infinity for stress states for which $\Phi\notin[0,1]$, indicated as white
zones in Fig.\ \ref{fig:yield}a. This is an inconvenience for numerical
implementation, as virtually any stress state can be encountered during iterative
solution of the incremental constitutive equations resulting, for instance, from
application of the return mapping algorithm.


One way to overcome this problem is to use a \lq squared version' of the yield
function (\ref{eq:F}) defined by
  \begin{equation}
    \label{eq:squared}
    F_2(\bfs{\sigma},\bfs{\eta}) = -f^2(p) + \frac{q^2}{g^2(\theta)} ,
  \end{equation}
which corresponds to the same BP yield surface and is defined for arbitrary stress,
but looses convexity,\footnote
 {\modified{We have thoroughly tried to reformulate the BP model to make the yield
  function globally smooth and convex, but this has not been possible.}}
as illustrated in Fig.\ \ref{fig:yield}b. Here, the squared function $f^2(p)$ is
defined for arbitrary $p$ by the first branch in Eq.~(\ref{eq:f}). The nonconvex
yield function shows, in the terminology introduced by Brannon and Leelavanichkul
\cite{BrannonLeelavanichkul10}, a \lq false elastic domain' so that, employing a
return mapping algorithm it may happen (for certain trial stresses) that convergence
is reached at a wrong stress state. Nonconvexity may also lead to divergence of the
corresponding iterative scheme.

\section{Implicit yield function in the $(p,q)$-space}
\label{sec:implicit}


In the special case of the BP yield function defined by Eq.~(\ref{eq:F}), one can
use the general formulation of the implicit yield function, as presented in
Section~\ref{sec:general}. However, as only the pressure-dependent part $f(p)$
specified by Eq.~(\ref{eq:f}) is inconvenient for numerical implementation, it is
simpler and thus numerically more efficient to apply the implicit yield function
formulation in the $(p,q)$-space, as shown below.

Compared to Section~\ref{sec:general}, a somewhat different presentation style is
adopted here, which is oriented towards computer implementation using an automatic
differentiation (AD) technique and \emph{AceGen}, an automatic code generation
system \cite{Korelc02,Korelc09}. Accordingly, the specific formulae, such as that in
Eq.\ (\ref{eq:dFs}) are not provided, as they are not needed, and the focus is on
indicating the actual dependencies and on defining the implicit derivatives of
$\varrho_0$.

\begin{figure}
  \centerline{\scalebox{0.9}{\input{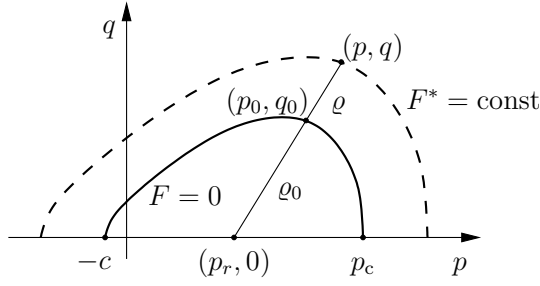}}}
  \caption{Construction of the implicit yield function $F^\ast$ in the $(p,q)$-space.
           \label{fig:implicit}}
\end{figure}

Consider the $(p,q)$-space and the yield surface $F=0$ corresponding to a fixed Lode
angle $\theta$, see Fig.~\ref{fig:implicit}. The implicit yield function $F^\ast$ is
given by, see Eq.~(\ref{eq:Fs:gen}),
  \begin{equation}
    \label{eq:Fs}
    F^\ast(\bfs{\sigma},\bfs{\eta}) = \frac{\varrho}{\varrho_0} - 1 ,
  \end{equation}
where $\varrho$ is now the distance between the current stress point $(p,q)$ in the
$(p,q)$--plane and a reference point $(p_r,0)$ taken inside the yield surface,
  \begin{equation}
    \varrho = \| \bfs{\varrho} \| , \quad
    \bfs{\varrho} = (p-p_r,q) ,
  \end{equation}
and $\varrho_0$ is the distance between the reference point $(p_r,0)$ and the image
$(p_0,q_0)$ of the current stress point $(p,q)$,
  \begin{equation}
    \label{eq:image:1}
    \varrho_0 = \| \bfs{\varrho}_0 \| , \quad
    \bfs{\varrho}_0 = (p_0-p_r,q_0) , \quad
    \bfs{\varrho} = (1+F^\ast) \bfs{\varrho}_0 .
  \end{equation}
The image point $(p_0,q_0)$, which is the counterpart of $\bfs{\sigma}_0$ of
Section~\ref{sec:general}, lies on the yield surface, thus
  \begin{equation}
    \label{eq:image:2}
    \hat{F}(p_0,q_0,\theta,\bfs{\eta}) = 0 ,
  \end{equation}
$\hat{F}$ being the yield function expressed in terms of the invariants, cf.\
Eq.~(\ref{eq:F}). In the present implementation, the squared version of the yield
function, Eq.~(\ref{eq:squared}), is actually used instead of the original version
(\ref{eq:F}), as the former is more convenient for practical application.

By construction, the yield function $F^\ast$ defined by Eq.~(\ref{eq:Fs}) generates
a family of self-similar surfaces $F^\ast=\mbox{const}$ that are scaled with respect
to the reference point $(p_r,0)$, as Fig.~\ref{fig:implicit} illustrates.
Self-similarity implies that the implicit yield function is convex if the generating
yield surface $F=0$ is convex.

\begin{figure}[!htcb]
  \centerline{
    \begin{tabular}{l}
      (a) 
      \inclps{!}{0.38\textwidth}{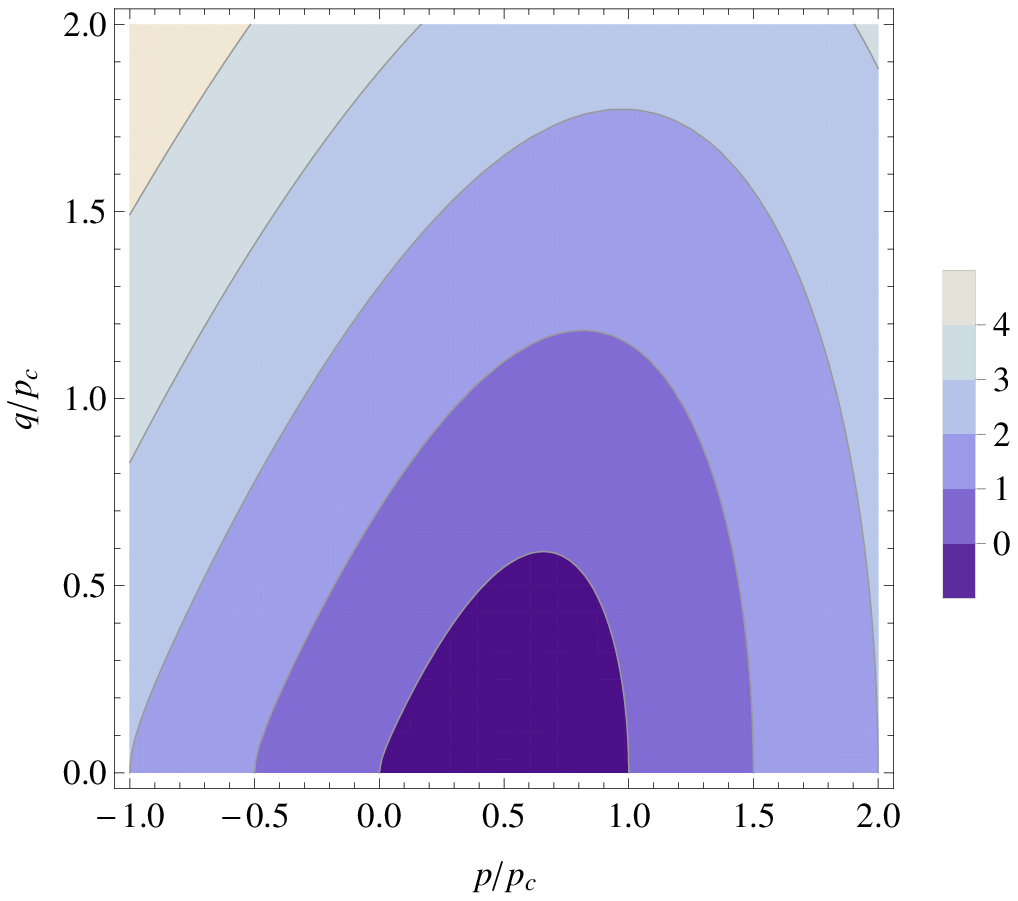} 
      (b) 
      \inclps{0.43\textwidth}{!}{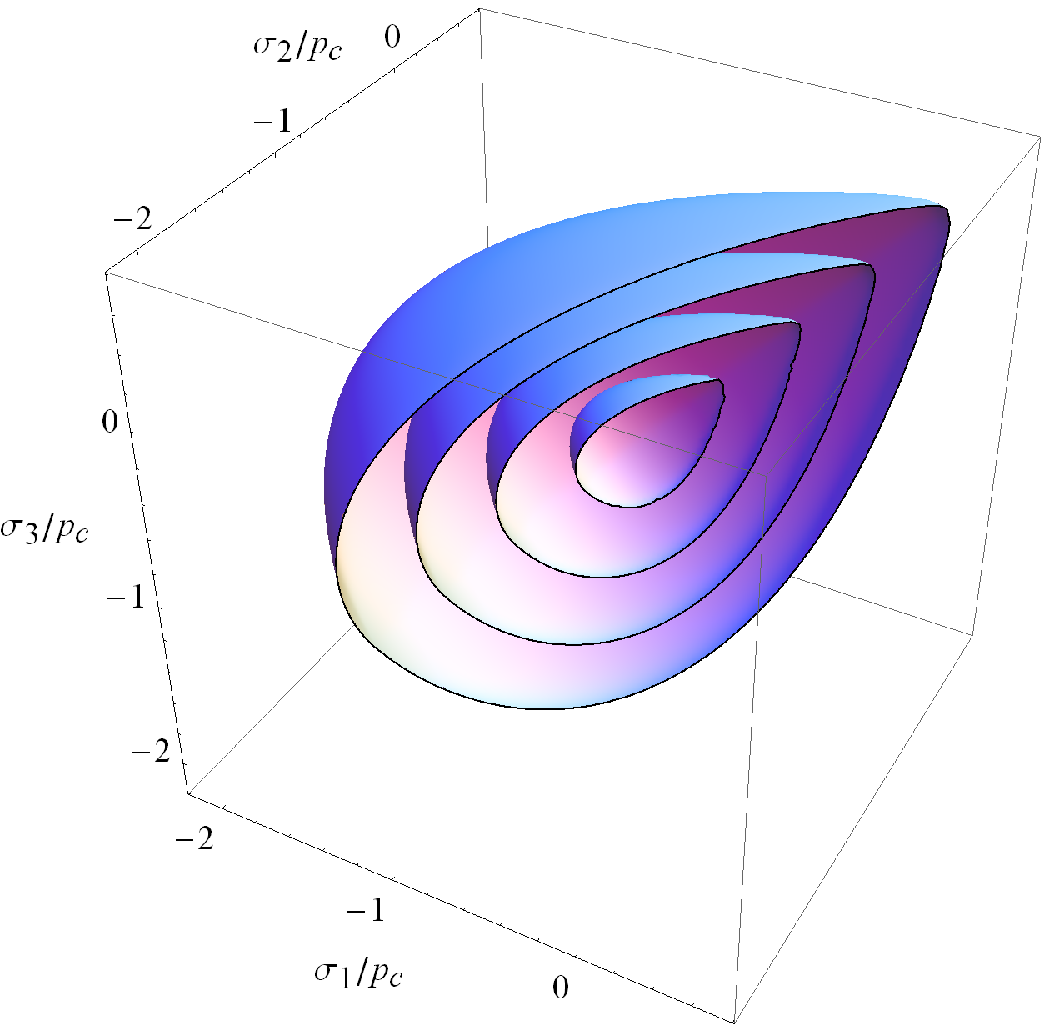}
    \end{tabular}
    }
  \caption{The iso-lines (iso-surfaces) of the implicit BP yield function:
           (a) in the $(p,q)$-space, (b) in the principal stress space.
           \label{fig:implicit:2}}
\end{figure}

The iso-lines of the implicit BP yield function
in the $(p,q)$-plane are shown in Fig.~\ref{fig:implicit:2}a (the model parameters
are equal to those adopted in Fig.~\ref{fig:yield}). Figure~\ref{fig:implicit:2}b
shows the iso-surfaces of the implicit BP yield function in the principal stress
space. Here and in the following, the position of the reference point has been
assumed as $p_r=(p_c+c)/2$, although other choices (for instance, the center of mass
of the yield surface \cite{Penasa14}) can be made.

The value of the yield function $F^\ast$ is defined implicitly by Eq.\
(\ref{eq:image:2}), i.e., by the condition that the image point lies on the yield
surface. This can be formulated as a nonlinear equation for $\varrho_0$ (at fixed
$\bfs{\sigma}$ and $\bfs{\eta}$),
  \begin{equation}
    \label{eq:F0}
    F_0 ( \varrho_0, \bfs{\sigma}, \bfs{\eta} ) = 0 ,
  \end{equation}
which is solved using the iterative Newton method,
  \begin{equation}
    \varrho_0^{i+1} = \varrho_0^i + \Delta \varrho_0^i , \quad
    \Delta \varrho_0^i = - \left( \frac{\partial F_0}{\partial \varrho_0} \right)^{-1}
      F_0 (\varrho_0^i) .
  \end{equation}

The solution $\varrho_0$ implicitly depends on the stress $\bfs{\sigma}$ and
hardening variables $\bfs{\eta}$ through Eq.\ (\ref{eq:F0}). The derivative of
$\varrho_0$ with respect to $\bfs{\sigma}$ is obtained by taking the total
derivative of Eq.\ (\ref{eq:F0}) with respect to $\bfs{\sigma}$,
  \begin{equation}
    \label{eq:dF0}
    \frac{\partial F_0}{\partial \varrho_0} \frac{\partial \varrho_0}{\partial\bfs{\sigma}}
      + \frac{\partial F_0}{\partial\bfs{\sigma}} = 0 ,
  \end{equation}
so that
  \begin{equation}
    \label{eq:dr0}
    \frac{\partial \varrho_0}{\partial\bfs{\sigma}} = - \left(
      \frac{\partial F_0}{\partial \varrho_0} \right)^{-1}
      \frac{\partial F_0}{\partial\bfs{\sigma}} .
  \end{equation}
The second derivative,
which is also needed, is obtained by differentiating (\ref{eq:dF0}) with respect to
$\bfs{\sigma}$ again, which gives
  \begin{equation}
  \label{eq:d2r0}
    \frac{\partial^2 \varrho_0}{\partial\bfs{\sigma}\partial\bfs{\sigma}} =
      - \left( \frac{\partial F_0}{\partial \varrho_0} \right)^{-1}
      \Bigg(
      \frac{\partial^2 F_0}{\partial \varrho_0\partial \varrho_0}
        \frac{\partial \varrho_0}{\partial\bfs{\sigma}} \otimes
        \frac{\partial \varrho_0}{\partial\bfs{\sigma}}
      + \frac{\partial^2 F_0}{\partial\bfs{\sigma}\partial\bfs{\sigma}}
      + 2 \left( \frac{\partial^2 F_0}{\partial\bfs{\sigma}\partial \varrho_0}
        \otimes \frac{\partial \varrho_0}{\partial\bfs{\sigma}}
      \right)_{\it sym} \Bigg) ,
  \end{equation}
Derivatives of $\varrho_0$ with respect to hardening variables $\bfs{\eta}$, as well
as mixed second derivatives, are obtained analogously. In fact, the formulae for the
implicit derivatives of $\varrho_0$ with respect to $\bfs{\eta}$ are obtained by
simply replacing $\bfs{\sigma}$ by $\bfs{\eta}$ in Eqs.~(\ref{eq:dr0}) and
(\ref{eq:d2r0}), with the adequate redefinition of the tensor product in
Eq.~(\ref{eq:d2r0}).

The first derivative $\partial \varrho_0/\partial\bfs{\sigma}$, Eq.~(\ref{eq:dr0}),
is needed, for instance, to compute the gradient of the implicit yield function, see
Eq.~(\ref{eq:dFs1}). The second derivative comes into play when the constitutive
update problem is solved and when the incremental constitutive equations are
linearized, see Section~\ref{sec:return}.



Computer implementation of the above implicit BP yield function has been performed
using \emph{AceGen}, a code generation system that combines the symbolic algebra
capabilities of \emph{Mathematica} (www.wolfram.com) with an automatic
differentiation technique and a stochastic expression optimization technique
\cite{Korelc02,Korelc09}. In particular, all the necessary explicit derivatives are
directly derived by automatic differentiation, while the implicit derivatives of
$\varrho_0$, i.e., those specified by Eqs.~(\ref{eq:dr0}) and (\ref{eq:d2r0}), are
introduced using the so-called AD exceptions implemented in \emph{AceGen}
\cite{Korelc09}.

The \emph{AceGen} implementation of the implicit BP yield function is provided as a
supplementary material accompanying this paper. Specifically, the complete
\emph{AceGen} input is provided for generating the numerical code that computes the
implicit BP yield function $F^\ast$ and its first and second derivatives with
respect to the stress $\bfs{\sigma}$ and with respect to all the parameters defining
the yield surface. The \emph{AceGen} code, as well as the corresponding ready-to-use
subroutines in \emph{C}, \emph{Fortran}, \emph{Mathematica} and \emph{Matlab}, are
also available at our web sites.\footnote
  {\url{http://www.ippt.pan.pl/~sstupkie/files/BPyield.html},
   \url{http://ssmg.unitn.it/BPyield.html}}

\section{Application to return mapping algorithm}
\label{sec:examples}

In this section, feasibility of the implicit yield function concept is illustrated
in the context of time integration algorithms in elastoplasticity. The classical
return mapping algorithm is first recalled and its performance is subsequently
evaluated in the case when the implicit BP yield function is used in an
elastoplastic model based on perfect plasticity.


\subsection{Return mapping algorithm}
\label{sec:return}

The return mapping algorithm is here introduced for the possibly simplest model of
elastoplasticity. Specifically, ideal plasticity (no hardening) and the associated
flow rule are assumed. No restrictions are introduced for the yield surface, except
the usual assumptions of convexity and smoothness. The goal is to apply the return
mapping algorithm to a yield surface that is defined by an implicit yield function,
such as the BP yield surface, and this can be sufficiently accomplished for the
present simple model. For more general formulations of elastoplasticity and their
computational treatment, the reader is referred, for instance, to the monographs
\cite{SimoHughes98,SouzaNeto08}.

Upon backward-Euler integration of the flow rule, the incremental constitutive
equations can be written in the form of the following constitutive update problem:

\vspace{2ex}

\noindent {\em {\bf Constitutive update problem:} Given the strain
$\bfs{\varepsilon}_{n+1}$ at the current time step $t=t_{n+1}$ and the plastic
strain $\bfs{\varepsilon}^p_n$ at the previous time step $t=t_n$, find the current
plastic strain $\bfs{\varepsilon}^p_{n+1}$ and the plastic multiplier  $\Delta\gamma$
that satisfy the elastic constitutive equation
  \begin{equation}
    \label{eq:sigma}
    \bfs{\sigma}_{n+1} = \mathbbm{C} [ \bfs{\varepsilon}_{n+1} - \bfs{\varepsilon}^p_{n+1} ] ,
  \end{equation}
the incremental flow rule
  \begin{equation}
    \label{eq:flow}
    \bfs{\varepsilon}^p_{n+1} = \bfs{\varepsilon}^p_n +
      \Delta\gamma \, \bfm{n}(\bfs{\sigma}_{n+1}) , \quad
    \bfm{n} = \frac{\partial F}{\partial\bfs{\sigma}} ,
  \end{equation}
and the complementarity conditions
  \begin{equation}
    \label{eq:compl}
    F(\bfs{\sigma}_{n+1}) \leq 0 , \quad
    \Delta\gamma \geq 0 , \quad
    \Delta\gamma F(\bfs{\sigma}_{n+1}) = 0 ,
  \end{equation}
where $F(\bfs{\sigma})$ is a given yield function with a convex and smooth zero
level set $F(\bfs{\sigma})=0$, and $\mathbbm{C}$ denotes the fourth-order elastic
moduli tensor.}

\vspace{2ex}

The constitutive update problem is solved using the return mapping algorithm which
involves the following steps:
\begin{enumerate}
\item Compute the trial elastic state
  \begin{equation}
    \bfs{\sigma}_{n+1}^{\it trial} =
      \mathbbm{C} [ \bfs{\varepsilon}_{n+1} - \bfs{\varepsilon}^p_n ] , \quad
			F^{\it trial} = F(\bfs{\sigma}_{n+1}^{\it trial}) .
  \end{equation}
\item Check the yield condition: if $F^{\it trial}\leq0$ then the step is
    elastic and
  \begin{equation}
    \bfs{\varepsilon}^p_{n+1} = \bfs{\varepsilon}^p_n , \quad \Delta\gamma = 0 .
  \end{equation}
\item If $F^{\it trial}>0$ then the step is plastic and the following nonlinear
    algebraic equations are solved for $\bfs{\varepsilon}^p_{n+1}$ and
    $\Delta\gamma$:
  \begin{equation}
    \label{eq:return}
    \begin{array}{l}
      0 = \bfs{\varepsilon}^p_{n+1} - \bfs{\varepsilon}^p_n -
        \Delta\gamma \, \bfm{n}(\bfs{\sigma}_{n+1}) , \\[0.5ex]
      0 = F(\bfs{\sigma}_{n+1}) ,
    \end{array}
  \end{equation}
  where the stress $\bfs{\sigma}_{n+1}$ is given by the constitutive
  equation~(\ref{eq:sigma}).
\end{enumerate}

Denote the vector of unknowns and the residual vector corresponding to the nonlinear
equations (\ref{eq:return}) by $\brm{h}=\{\bfs{\varepsilon}^p_{n+1},\Delta\gamma\}$
and $\brm{Q}(\brm{h})$, respectively. Equation $\brm{Q}(\brm{h})=\brm{0}$ is then
solved using the Newton method which employs the following iterative scheme:
  \begin{equation}
    \label{eq:Newton}
    \brm{h}^{i+1} = \brm{h}^i + \Delta\brm{h}^i , \quad
    \Delta\brm{h}^i = - \left( \frac{\partial\brm{Q}}{\partial\brm{h}} \right)^{-1}
      \brm{Q}(\brm{h}^i) ,
  \end{equation}
with the typical initial guess $\brm{h}^0=\{\bfs{\varepsilon}^p_n,0\}$.

At the solution of the constitutive update problem, the current stress
$\bfs{\sigma}_{n+1}$ belongs to the
elastic domain $F(\bfs{\sigma})\leq0$. However, the trial stress
$\bfs{\sigma}_{n+1}^{\it trial}$ and the stresses $\bfs{\sigma}_{n+1}^i$ at
subsequent iterations may lie well outside the elastic domain. It is thus crucial
for the direct application of the return mapping algorithm that the yield function
$F(\bfs{\sigma})$ be defined and differentiable for arbitrary stress. The original
BP yield function (but also many other yield functions, for instance, those of
Jeremic et al.~\cite{Jeremic99} and Foster et al.~\cite{Foster05}) is thus not
suitable for the return mapping algorithm, while the implicit one introduced in
Section~\ref{sec:implicit} is fully suitable.

Note that the flow rule (\ref{eq:flow}) involves the gradient of the yield function.
The tangent matrix $\partial\brm{Q}/\partial\brm{h}$ used in the Newton scheme
(\ref{eq:Newton}) involves thus the second derivative of the yield function. In the
case of the implicit yield function, those derivatives cannot be obtained directly,
so that the implicit derivatives discussed in Section~\ref{sec:implicit} must be
used instead.




Convergence of the iterative Newton scheme (\ref{eq:Newton}) is not guaranteed in
general. This difficulty can be circumvented (at least partially) by applying a
globally convergent scheme, for instance, by enhancing the Newton scheme with a
suitable line search strategy. The iterative update (\ref{eq:Newton})$_1$ would then
be replaced by the following one: $\brm{h}^{i+1}=\brm{h}^i+\alpha^i\Delta\brm{h}^i$,
where the line search parameter $0<\alpha^i\leq1$ is determined by requiring that
the iterative update results in a sufficient decrease of an appropriate merit
function, see, for instance, \cite{NocedalWright06}. In the numerical study reported
in Section~\ref{sec:study}, the Newton method combined with the line search scheme
specified in Box~A.1 of~\cite{PerezFoguetArmero02} has been used in addition to the
pure Newton method specified by Eq.~(\ref{eq:Newton}).
More elaborate, globally convergent schemes employing primal and dual algorithms for
the solution of the closest-point projection problems in incremental
elastoplasticity are discussed in \cite{PerezFoguetArmero02}.


\subsection{Implementation and computational efficiency}
\label{sec:efficiency}

For the purpose of
testing of the proposed implicit yield function formulation, a simple computer code
has been developed that implements the return mapping algorithm described in the
previous section. The code solves the constitutive update problem
(\ref{eq:sigma})--(\ref{eq:compl}) in a format that exactly corresponds to the
constitutive model implementation in a displacement-based finite element code.
Specifically, for a given total strain $\bfs{\varepsilon}_{n+1}$ and previous
plastic strain $\bfs{\varepsilon}^p_n$, the current plastic strain
$\bfs{\varepsilon}^p_{n+1}$ is computed along with the current stress
$\bfs{\sigma}_{n+1}$ and consistent tangent moduli
  \begin{equation}
    \mathbbm{C}^{\it ep}_{n+1} =
      \frac{\partial\bfs{\sigma}_{n+1}}{\partial\bfs{\varepsilon}_{n+1}} .
  \end{equation}


While the code developed to test the proposed formulation is limited for simplicity
to material-point computations and small-strain ideal elastoplasticity, the implicit
yield function formulation has already been successfully implemented in a
finite-element framework and for a much more general class of constitutive models.
In particular, the model for ceramic powder compaction
\cite{Piccolroaz06-1,Piccolroaz06-2} has been implemented in its finite-strain
version including effects such as nonlinear hardening, elastoplastic coupling, and
non-associated flow rule. Some examples of the corresponding finite-element
simulations of ceramic powder forming processes, but without details of algorithmic
treatment, are reported in \cite{StupPicBig14}.


Computer implementation of the incremental constitutive equations has been performed
using the \emph{AceGen} system \cite{Korelc02,Korelc09}. The formulation of
incremental elastoplasticity that is appropriate for an automated implementation
using \emph{AceGen} is introduced in \cite{Korelc09}; a concise presentation of the
formulation can also be found in \cite{KorelcStupkiewicz14}. As an essential part of
this formulation, an automatic differentiation technique is used to automatically
derive the consistent tangent moduli $\mathbbm{C}^{\it ep}_{n+1}$ and other relevant
quantities such as the dependent tangent $\partial\brm{Q}/\partial\brm{h}$ involved
in the Newton scheme (\ref{eq:Newton}).
In the context of the present implicit yield function formulation, the automatic
differentiation technique
is also used to efficiently implement the implicit derivatives (\ref{eq:dr0}) and
(\ref{eq:d2r0}), representing an important \lq ingredient' in the formulation. This
part of \emph{AceGen} implementation is available as a supplementary material, see
Section~\ref{sec:implicit}.

It is obvious that the use of an implicitly defined yield function, as introduced in
Section~\ref{sec:implicit}, in incremental algorithms of elastoplasticity is
necessarily associated with an extra computational cost due to the nonlinear
equation (\ref{eq:F0}), that must be solved at each evaluation of the yield
function. Furthermore, evaluation of the gradient and second derivative of the yield
function requires evaluation of the implicit derivatives (\ref{eq:dr0}) and
(\ref{eq:d2r0}), and this also has a computational price. Therefore, it is
worthwhile to assess the computational efficiency of the implicit yield function
formulation, and this is done below with reference to the Cam-clay model.

The original Cam-clay yield function is defined as $F_{cc}=(q/M)^2+p(p-p_c)$, but
instead than this, the transformed Cam-clay yield function $F_{cc}^\ast$ will now be
used,
  \begin{equation}
    F_{cc}^\ast(\bfs{\sigma}) = \sqrt{ \left( \frac{2q}{M p_c} \right)^2 +
      \left( \frac{2p}{p_c} -1 \right)^2 } - 1 ,
  \end{equation}
which specifies the same yield surface $F_{cc}^\ast=0$ as the original Cam-clay
yield function, $F_{cc}=0$. At the same time, the transformed yield function
$F_{cc}^\ast$ is equivalent to the implicit BP yield function $F^\ast$,
Eq.~(\ref{eq:Fs}), obtained in the special case corresponding to the Cam-clay yield
surface, by setting $c=0$, $m=2$, $\alpha=1$, $\beta=1$, and $\gamma=0$,
cf.~\cite{BigoniPiccolroaz04}.

Being numerically equivalent, the two yield functions exhibit identical convergence
behavior in the return mapping algorithm, but they differ in computational
efficiency: $F_{cc}^\ast$ is an explicit analytical function while $F^\ast$ is
implicit and thus is associated with an additional computational cost, as described
above.



Table~\ref{tab:efficiency} presents the results of a study of computational
efficiency of the implicit BP yield function $F^\ast$. The constitutive update
problem has been repeatedly solved for several values of input variables, and the
corresponding overall evaluation time has been determined. That procedure has been
applied for the Cam-clay yield function $F_{cc}^\ast$ and for the implicit BP yield
function $F^\ast$, with parameters adjusted such that the response of the two is
identical. The specific model parameters ($E$, $\nu$, $p_c$, $M$) used in the
present study are given in Table~\ref{tab:parameters} (and are typical of alumina
powder), while the remaining parameters of the BP yield function assume the special
values that define the Cam-clay yield surface (see above).

\begin{table*}
  \caption{Normalized evaluation time and code size for the solution of the constitutive
           update problem for the Cam-clay yield function.
           \label{tab:efficiency}}
  \vspace{2ex}
  \centerline{
    \renewcommand{\arraystretch}{1.25}
    \begin{tabular}{lcccccc} \hline
      & & \multicolumn{2}{c}{Cam-clay yield function $F_{cc}^\ast$}
      & & \multicolumn{2}{c}{Implicit BP yield function $F^\ast$} \\ \cline{3-4} \cline{6-7}
      & & ~~ $\bfs{\sigma}_{n+1}$ ~~ & $\bfs{\sigma}_{n+1}$ \& $\mathbbm{C}^{\it ep}_{n+1}$
      & & ~~ $\bfs{\sigma}_{n+1}$ ~~ & $\bfs{\sigma}_{n+1}$ \& $\mathbbm{C}^{\it ep}_{n+1}$ \\ \hline
      Evaluation time & & 1.000 & 1.032 & & 4.148 & 4.408 \\
      Code size       & & 1.000 & 1.506 & & 2.916 & 3.824 \\ \hline
    \end{tabular}
    }
\end{table*}

In addition to the computation of both the stress and the consistent tangent moduli
(marked ``$\bfs{\sigma}_{n+1}$ \& $\mathbbm{C}^{\it ep}_{n+1}$'' in
Table~\ref{tab:efficiency}), the code has also been tested when computing only the
stress (marked ``$\bfs{\sigma}_{n+1}$'' in Table~\ref{tab:efficiency}). The
evaluation times, normalized by the evaluation time needed for the computation of
the stress using the Cam-clay yield function $F_{cc}^\ast$, are reported in
Table~\ref{tab:efficiency}.

The return mapping algorithm based on the implicit yield function $F^\ast$ has been
found to be approximately four times less efficient than that employing the explicit
Cam-clay yield function $F_{cc}^\ast$. It is reminded here that the convergence
behavior, including the number of Newton iterations, of both formulations is
identical, so that the difference is solely due to the extra cost of evaluation of
the implicit yield function and its derivatives. Though the factor of four might at
a first glance seem significant, two aspects have to be considered. Firstly, a very
simple elastoplastic model has been used in the present study, so that the
evaluation of the yield function and its derivatives constitutes a significant part
of the complete formulation. In more complex models (involving, for instance,
hardening, elastoplastic coupling, or finite-strain effects), the extra cost of
evaluation of the implicit yield function related to the (increased) overall cost of
the constitutive update problem becomes significantly lower. Secondly, the
remarkable flexibility of the meridian and deviatoric shape of the BP yield surface
must be associated to some \lq computational cost'.

Interestingly, the additional computing cost for the consistent tangent moduli is
surprisingly small, when compared to the computing cost of the stress only, as it is
well below 10\% for both formulations. This is probably due to the highly efficient
treatment of implicit dependencies through the so-called AD exceptions that are
implemented in \emph{AceGen}, see~\cite{Korelc09}.




Table~\ref{tab:efficiency} shows also the normalized size of the generated C code.
As expected, the use of the implicit yield function $F^\ast$ results in an increase
of the code size by a factor of 2--3, with respect to the code corresponding to the
explicit yield function $F_{\rm cc}^\ast$. The code size is in some way related to
the evaluation time, so that the remarks concerning the latter apply also here.


\subsection{Convergence of the return mapping algorithm}
\label{sec:study}

\begin{figure}
  \centerline{\inclps{0.5\textwidth}{!}{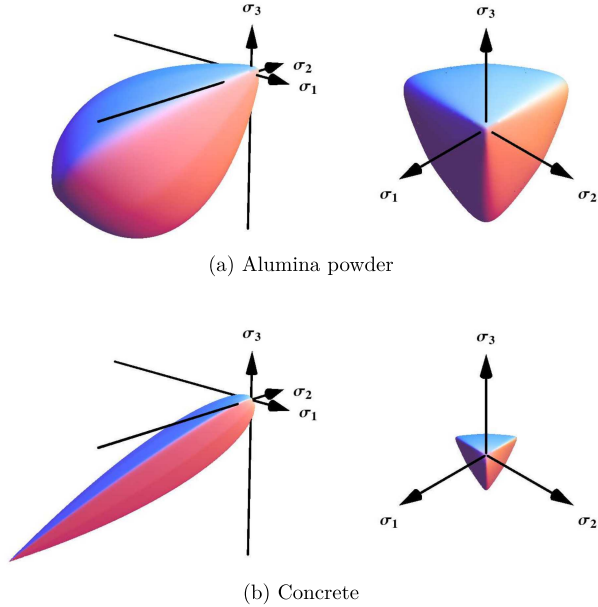}}
  \caption{Yield surfaces in the principal stress space: (a) alumina powder, (b) concrete.
           \label{fig:surfaces}}
\end{figure}

In this section, convergence of the return mapping algorithm is analyzed for two
sets of model parameters that correspond to alumina powder \cite{Piccolroaz06-1} and
concrete \cite{Penasa14}. The model parameters are provided in
Table~\ref{tab:parameters}, and the corresponding yield surfaces are shown in
Fig.~\ref{fig:surfaces}.

\begin{table*}
  \caption{Model parameters used in the convergence study.
           \label{tab:parameters}}
  \vspace{2ex}
  \centerline{
    \renewcommand{\arraystretch}{1.25}
    \begin{tabular}{lccccccccc} \hline
      & $E$ (MPa) & $\nu$ & $p_c$ (MPa) & $c$ (MPa) & $M$ & $m$ & $\alpha$ & $\beta$ & $\gamma$ \\ \hline
      Alumina powder &  1000 & 0.3  &  10 & 0 & 1.1  & 2 & 0.1  & 0.19 & 0.9  \\
      Concrete       & 11200 & 0.18 & 350 & 2 & 0.26 & 2 & 1.99 & 0.12 & 0.98 \\ \hline
    \end{tabular}
    }
\end{table*}

\begin{figure*}
  \centerline{
    \begin{tabular}{cccc}
      \inclps{0.29\textwidth}{!}{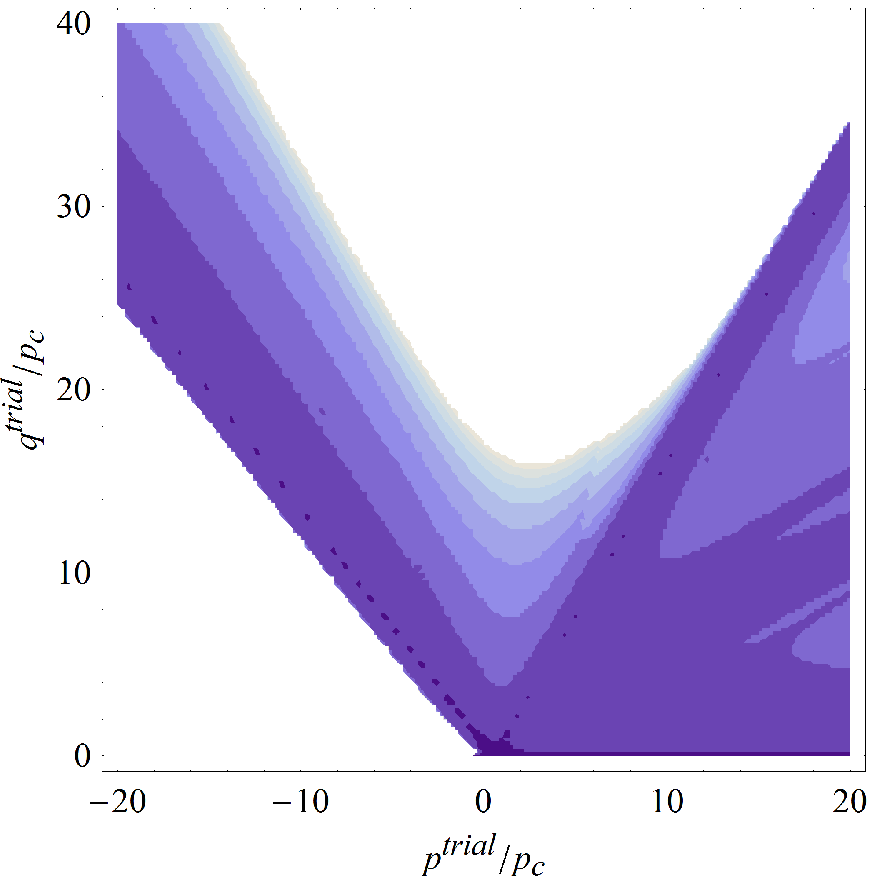} &
      \inclps{0.29\textwidth}{!}{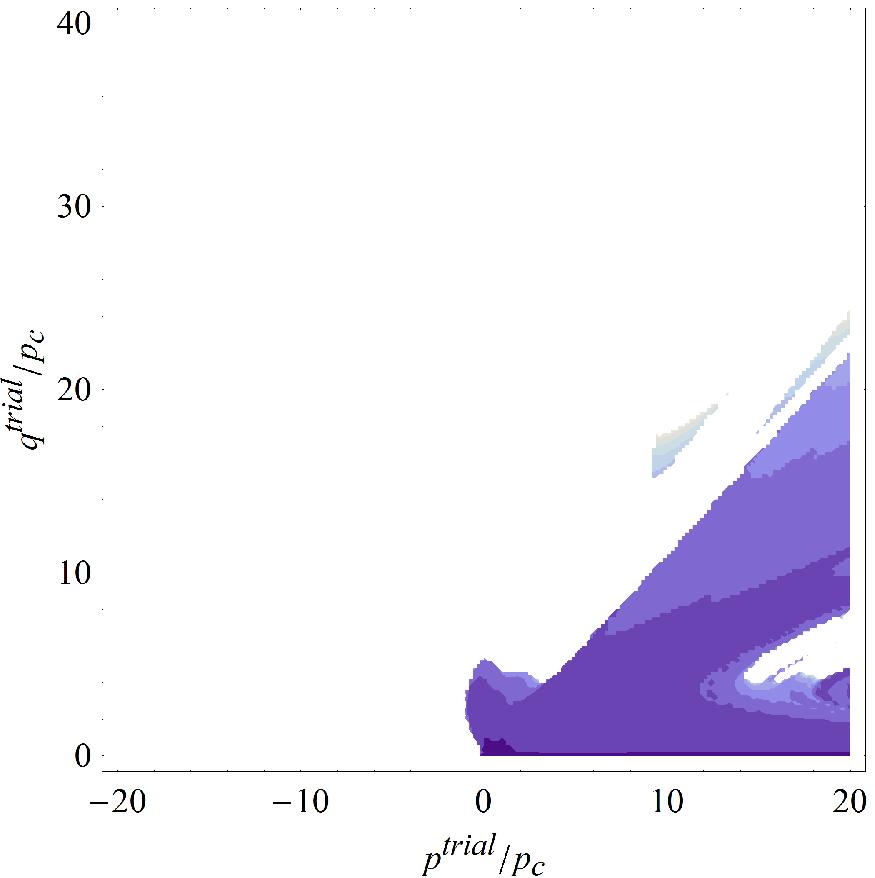} &
      \inclps{0.29\textwidth}{!}{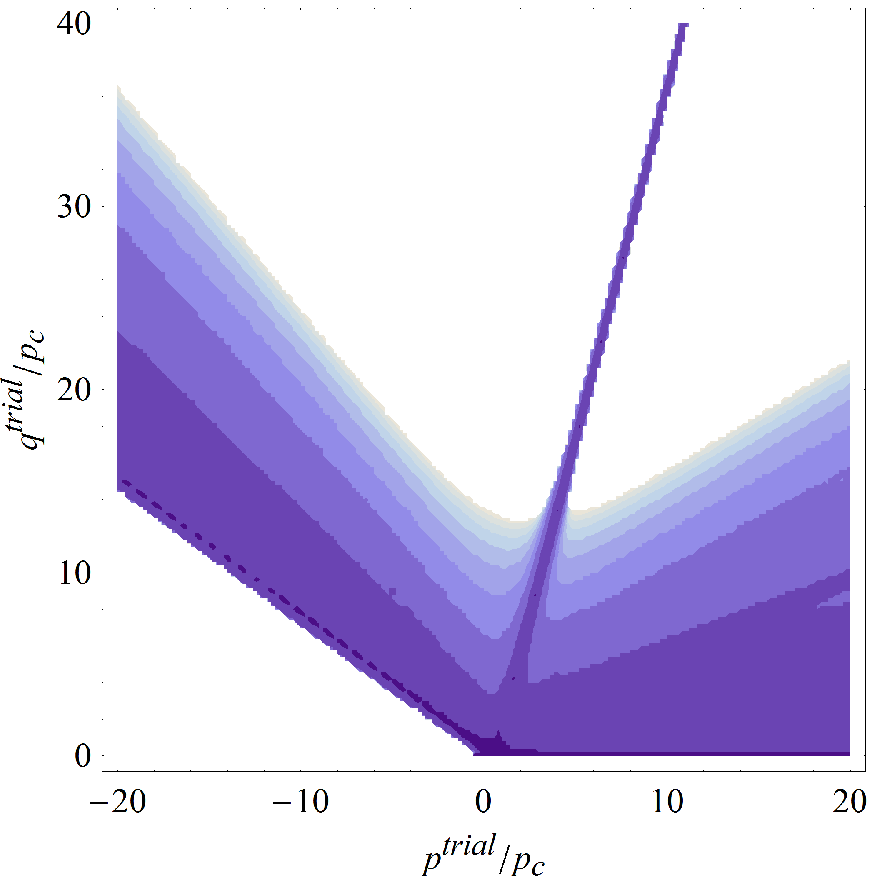} &
      \raisebox{3.2ex}{\inclps{!}{0.24\textwidth}{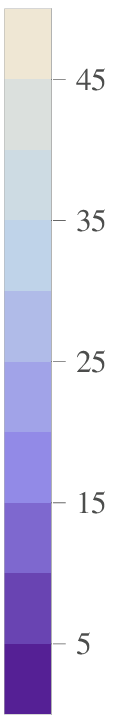}} \\[1ex]
      \inclps{0.29\textwidth}{!}{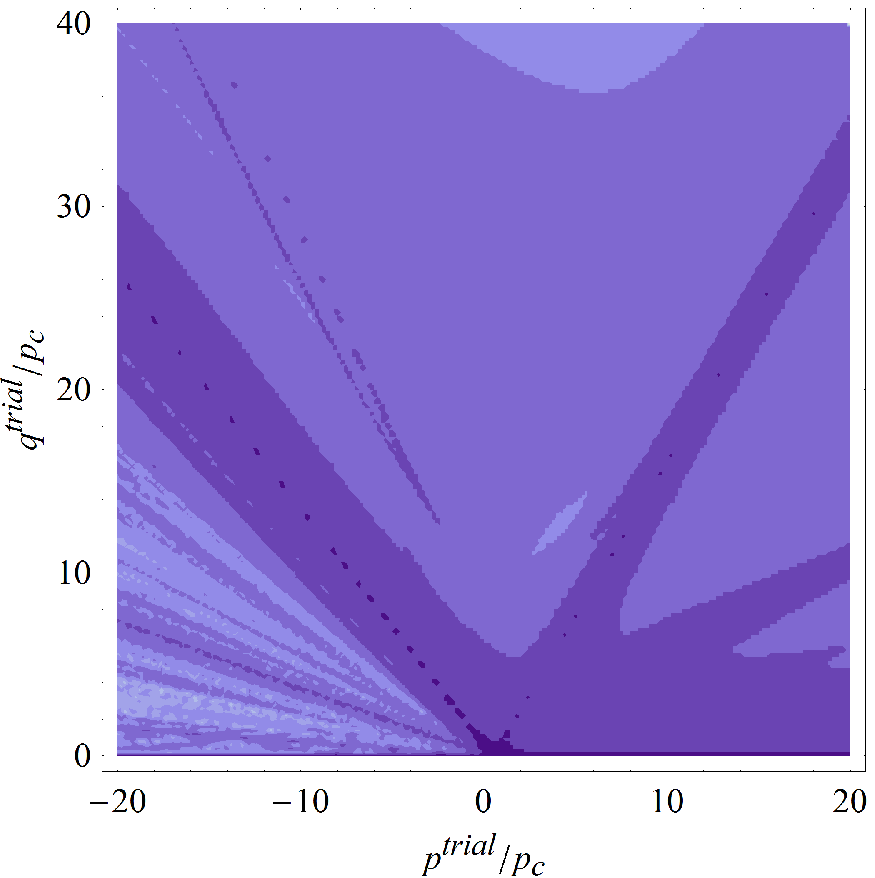} &
      \inclps{0.29\textwidth}{!}{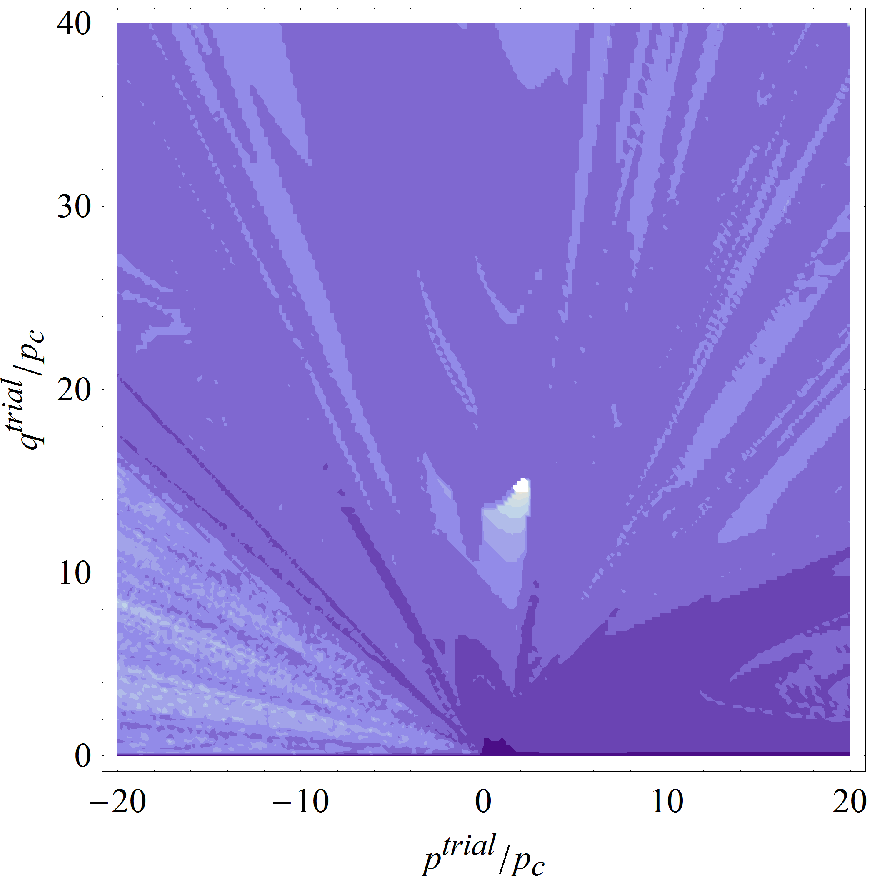} &
      \inclps{0.29\textwidth}{!}{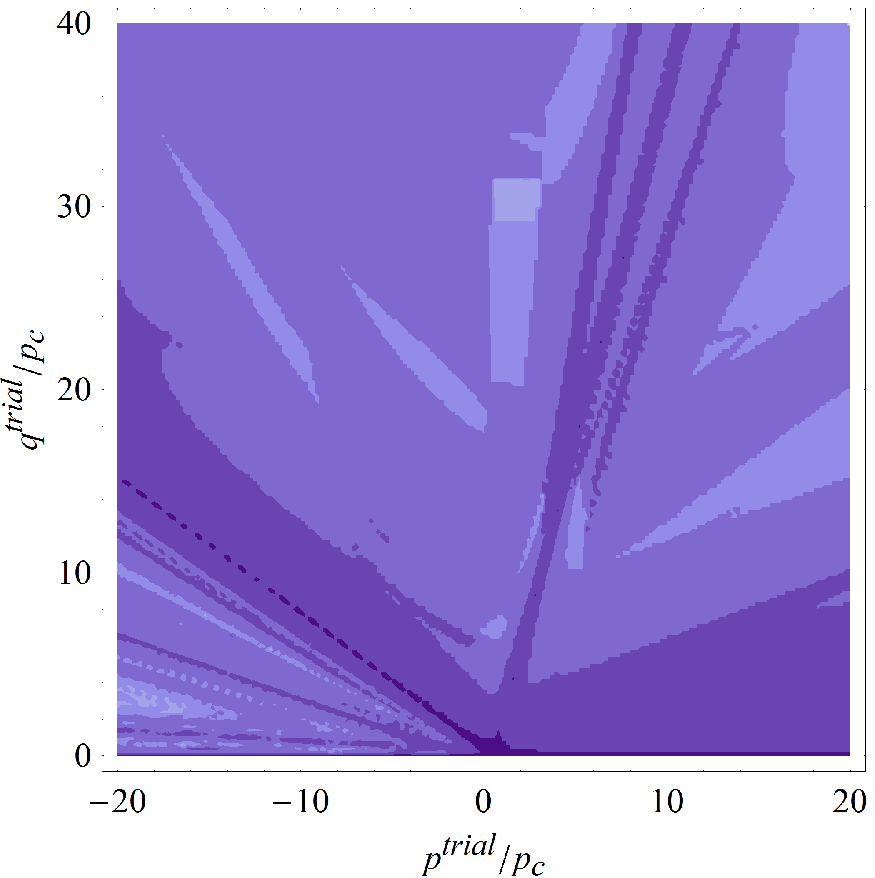} &
      \raisebox{3.2ex}{\inclps{!}{0.24\textwidth}{bp2_convergence_legend}} \\[0.5ex]
      ~~~(a) & ~~~(b) & ~~~(c) &
    \end{tabular}
    }
  \caption{Number of iterations needed for convergence of the return mapping algorithm
           using the Newton method (top row) and the Newton method with line search (bottom
           row) for (a) $\theta^{\it trial}=0$, (b) $\theta^{\it trial}=\pi/6$,
           (c) $\theta^{\it trial}=\pi/3$.
           White regions indicate lack of convergence (defined as more than 50 iterations).
           The parameters of the yield function correspond to alumina powder.
           \label{fig:conv:alumina}}
\end{figure*}

\begin{figure*}
  \centerline{
    \begin{tabular}{cccc}
      \inclps{0.29\textwidth}{!}{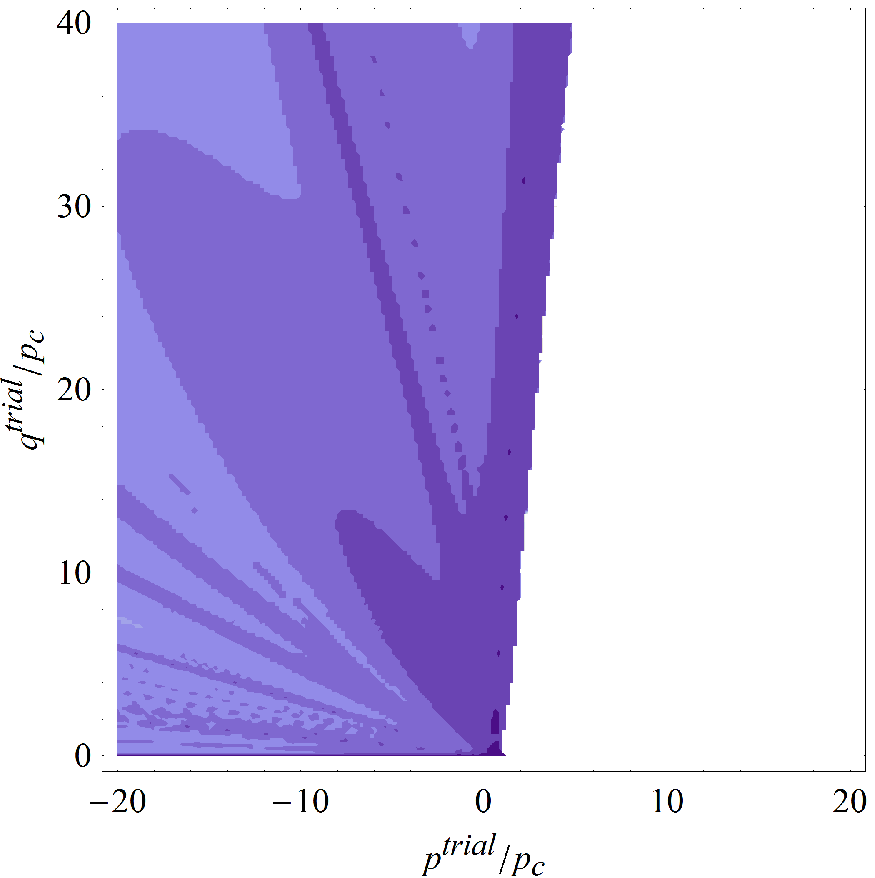} &
      \inclps{0.29\textwidth}{!}{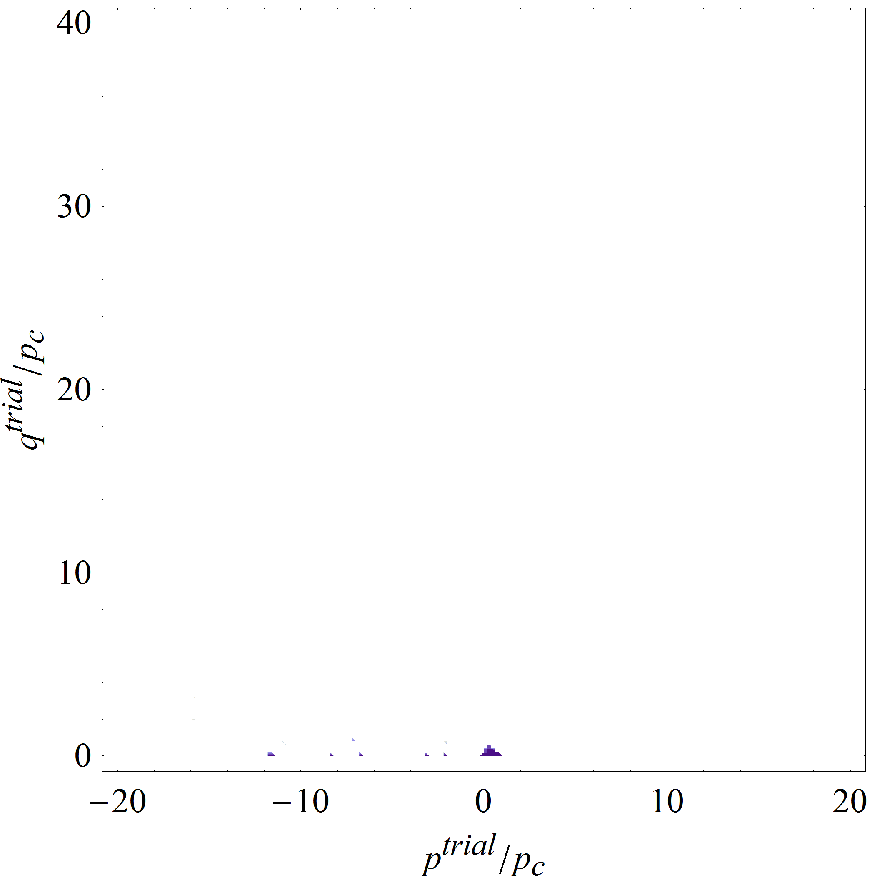} &
      \inclps{0.29\textwidth}{!}{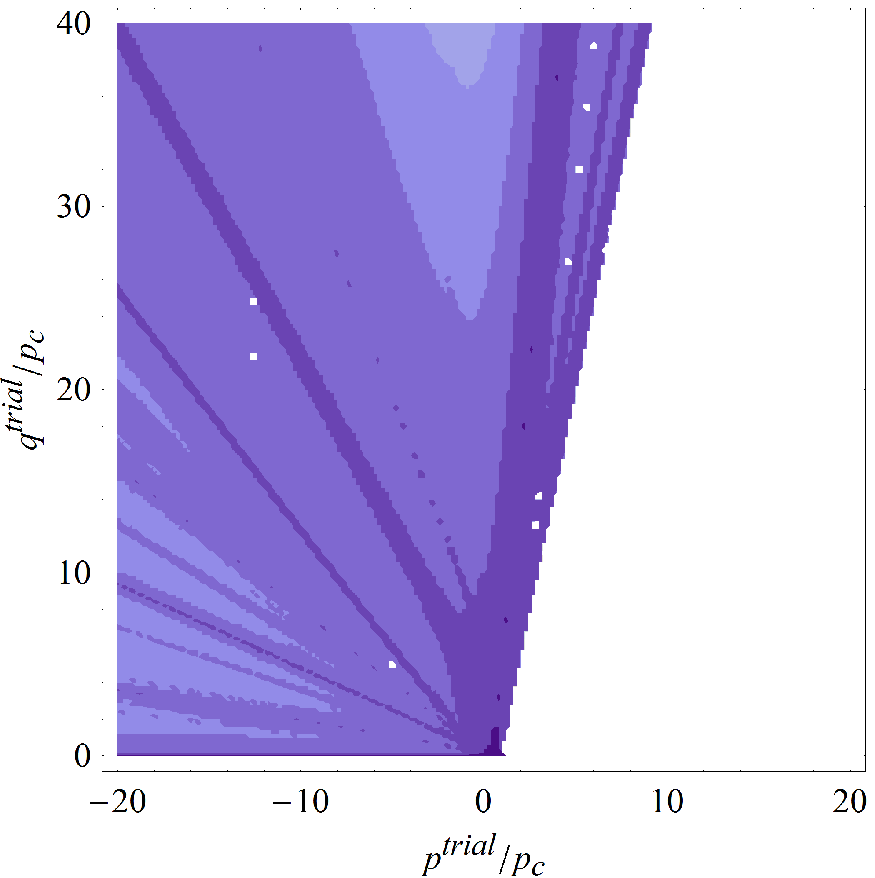} &
      \raisebox{3.2ex}{\inclps{!}{0.24\textwidth}{bp2_convergence_legend}} \\[1ex]
      \inclps{0.29\textwidth}{!}{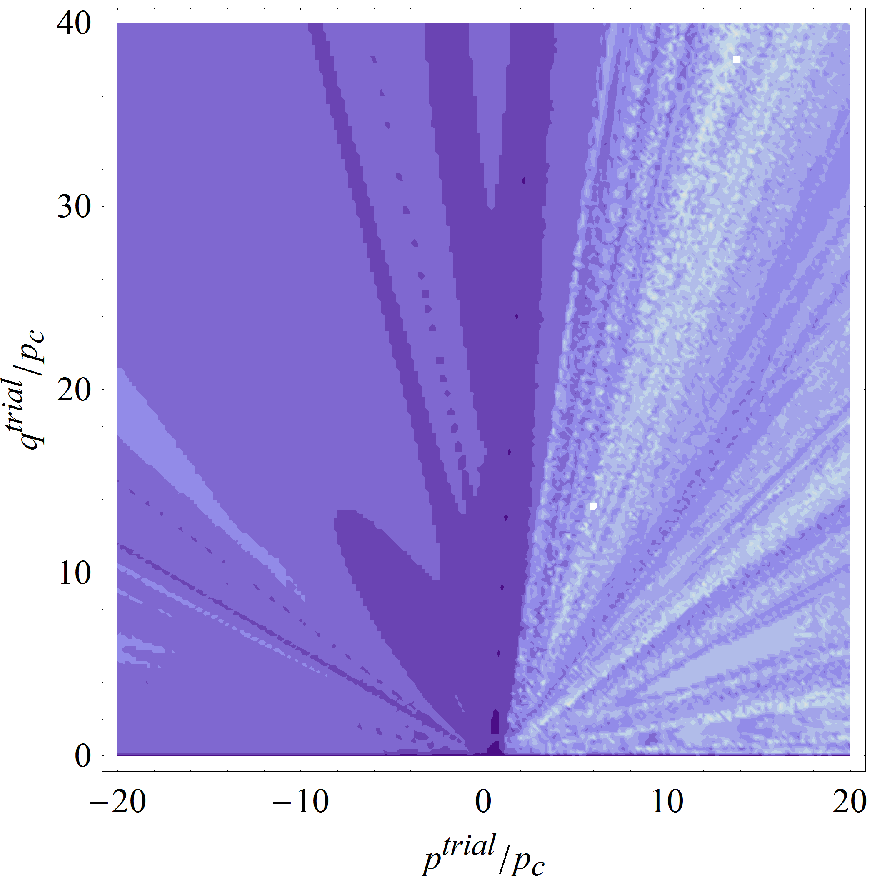} &
      \inclps{0.29\textwidth}{!}{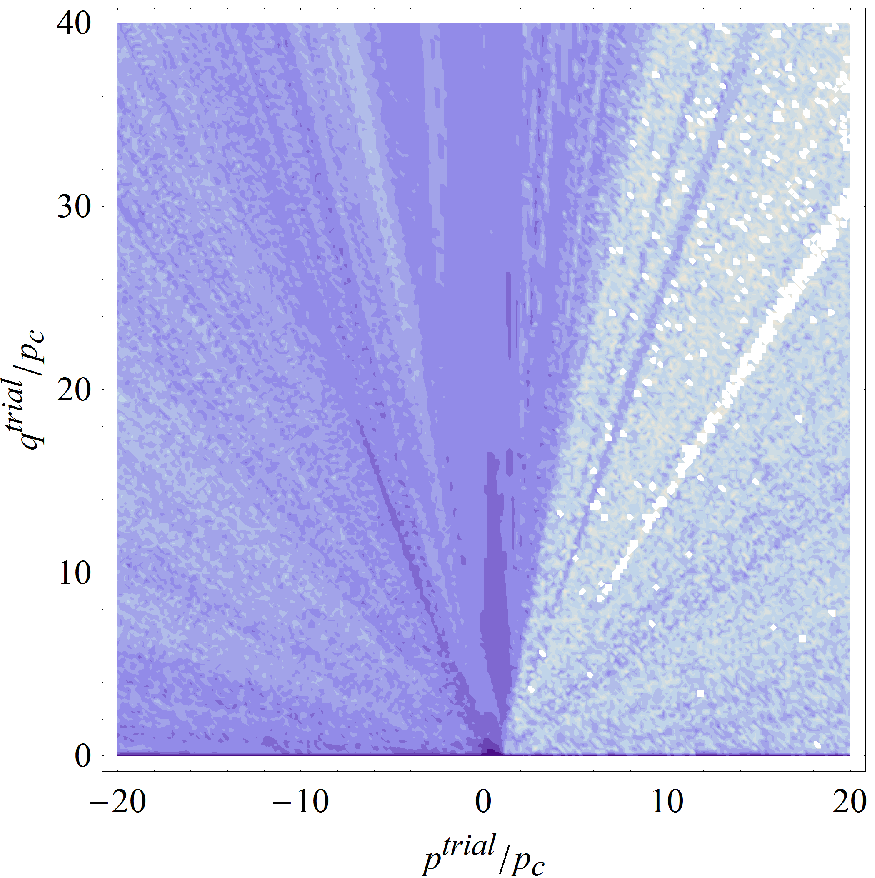} &
      \inclps{0.29\textwidth}{!}{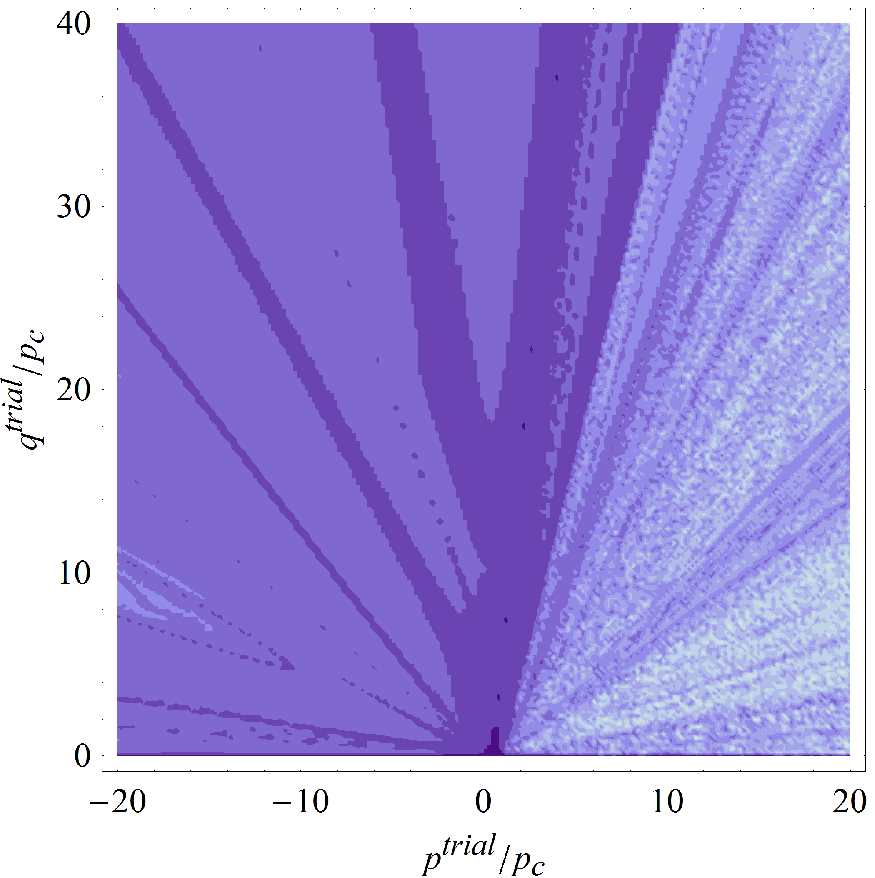} &
      \raisebox{3.2ex}{\inclps{!}{0.24\textwidth}{bp2_convergence_legend}} \\[0.5ex]
      ~~~{\small (a)} & ~~~{\small (b)} & ~~~{\small (c)} &
    \end{tabular}
    }
  \caption{Number of iterations needed for convergence of the return mapping algorithm
           using the Newton method (top row) and the Newton method with line search (bottom
           row) for (a) $\theta^{\it trial}=0$, (b) $\theta^{\it trial}=\pi/6$,
           (c) $\theta^{\it trial}=\pi/3$.
           White regions indicate lack of convergence (defined as more than 50 iterations).
           The parameters of the yield function correspond to concrete.
           \label{fig:conv:concrete}}
\end{figure*}

In the present study, the number of iterations needed for the convergence of the
return mapping algorithm has been evaluated as a function of the elastic trial
state. Due to isotropy, the trial stress $\bfs{\sigma}^{\it trial}$ is fully
characterized by its invariants $p^{\it trial}$, $q^{\it trial}$ and $\theta^{\it
trial}$, and Figs.~\ref{fig:conv:alumina} and~\ref{fig:conv:concrete} show the
number of iterations as a function of $p^{\it trial}$ and $q^{\it trial}$ for three
selected values of $\theta^{\it trial}$. Note that  $p^{\it trial}$ and $q^{\it
trial}$ have been normalized by $p_c$ so that the elastic range (in which the number
of iterations is obviously equal to zero) occupies a very small region corresponding
to $0\leq p^{\it trial}/p_c\leq1$ and $q^{\it trial}/p_c$ close to zero. Each
contour plot shown in Figs.~\ref{fig:conv:alumina} and~\ref{fig:conv:concrete} has
been obtained by sampling the trial $(p,q)$-space using 200$\times$200 points.

The maximum number of iterations has been set to 50, and the white regions in
Figs.~\ref{fig:conv:alumina} and~\ref{fig:conv:concrete} indicate the regions where
the maximum number of iterations has been reached and which are thus considered \lq
no-convergence regions'. It has been checked that, in the case of the pure Newton
method (reported in the top rows in Figs.~\ref{fig:conv:alumina}
and~\ref{fig:conv:concrete}), no-convergence regions do not noticeably decrease with
an increase of the maximum number of iterations. The figures corresponding to the
Newton method with line search (reported in the bottom rows in
Figs.~\ref{fig:conv:alumina} and~\ref{fig:conv:concrete}), show only isolated spots
of lack of convergence, and it has been checked that these regions vanish when the
maximum number of iterations is further increased.

The main purpose of the present numerical example is to show that the implicit yield
surface formulation can be effectively applied to develop incremental schemes for
elastoplasticity. This goal has been
achieved by showing that the implicit BP yield function can be evaluated for
arbitrary stresses, so that the standard return mapping algorithm can be applied
directly, while this is not the case of the original BP yield function, as
illustrated in Section~\ref{sec:BP}. Further, by construction, the implicit yield
function is convex (provided that the generating yield surface is convex), which is
beneficial for the convergence of the return mapping algorithm. This is illustrated
by the excellent convergence properties of the return mapping algorithm, employing the
Newton method combined with a
standard line search strategy.

Note that the two yield surfaces used in the present study pose difficulties in
their computational treatment as they locally exhibit very high curvature.
Especially the yield surface representative of concrete features sharp edges in the
deviatoric section ($\gamma=0.98$) and nearly a vertex on the hydrostatic axis
located at high pressure ($\alpha=1.99$). Those features are responsible for the
poor convergence of the pure Newton method, particularly for $\theta^{\it
trial}=\pi/6$, both for the model for alumina powder and for the model for concrete,
see Figs.~\ref{fig:conv:alumina}b and~\ref{fig:conv:concrete}b (in the latter case,
the no-convergence region spans the whole plot).

It should be noted here that the simple return mapping algorithm enhanced with a
line search strategy performs so well because the elastoplastic model is rather
simple, so that the only difficulty lies in the complexity of the yield surface, as
related, in particular, to its high curvature. Effects such as strain hardening and
elastoplastic coupling, which are crucial in applications involving granular and
rock-like materials, would definitely deteriorate convergence of the return mapping
algorithm. Anyway, the implicit yield function formulation may offer significant
advantages in the development of more advanced incremental formulations of
elastoplasticity.





\section{Conclusions}

A technique has been introduced to integrate elastoplastic (or elastodamaging) rate
equations based on pressure-sensitive and $J_3$-dependent yield functions that often
exhibit undesired features such as lack of convexity or \lq false elastic domains'.
The technique is based on a re-definition of the yield function in an implicit form,
that can be subsequently treated with standard numerical tools. Although associated
to a numerical cost, the proposed technique is shown to be efficient, general and
accurate.

\vspace*{10mm} \noindent {\sl Acknowledgments }
SS and DB gratefully acknowledge financial support from the European Union's Seventh
Framework Programme FP7/2007-2013/ under REA grant agreement number
PIAP-GA-2011-286110-INTERCER2. AP gratefully acknowledges financial support from the
European Union's Seventh Framework Programme FP7/2007-2013/ under REA grant
agreement number PITN-GA-2013-606878-CERMAT2.

\appendix
\section{Second derivative of $F^\ast$}
\label{app:second}

The second derivative of the implicit yield function is obtained by differentiating
Eq.~(\ref{eq:dFs1}), viz.
  \begin{equation}
\label{eq:d2Fs1}
  \frac{\partial^2 F^\ast}{\partial\bfs{\sigma}\partial\bfs{\sigma}} =
      \frac{1}{\varrho_0}
        \frac{\partial^2 \varrho}{\partial\bfs{\sigma}\partial\bfs{\sigma}}
      -\frac{\varrho}{\varrho_0^2}
        \frac{\partial^2 \varrho_0}{\partial\bfs{\sigma}\partial\bfs{\sigma}}
      +\frac{2\varrho}{\varrho_0^3} \frac{\partial \varrho_0}{\partial\bfs{\sigma}}
        \otimes \frac{\partial \varrho_0}{\partial\bfs{\sigma}}
      -\frac{2}{\varrho_0^2} \left( \frac{\partial \varrho}{\partial\bfs{\sigma}}
        \otimes \frac{\partial \varrho_0}{\partial\bfs{\sigma}} \right)_{\it sym} .
  \end{equation}
The above formula involves the explicit first and second derivatives of $\varrho$
and the implicit first and second derivatives of $\varrho_0$. The first derivative
of $\varrho$ is given by Eq.~(\ref{eq:drho}), the second derivative is given by
  \begin{equation}
    \frac{\partial^2 \varrho}{\partial\bfs{\sigma}\partial\bfs{\sigma}} =
      \frac{1}{\varrho} ( \mathbbm{S} - \bfm{e}_r \otimes \bfm{e}_r ) ,
  \end{equation}
in view of Eq.~(\ref{eq:der}). The implicit first derivative of $\varrho_0$ has been
derived in Section~\ref{sec:general}, see Eq.~(\ref{eq:drho0}), and the second
derivative is obtained below.

For future use, we introduce the following notation for the derivatives of the
original yield function evaluated at $\bfs{\sigma}_0$,
  \begin{equation}
    \bfm{n}_0 =
      \left. \frac{\partial F}{\partial\bfs{\sigma}} \right|_{\bfs{\sigma}_0} , \quad
    \mathbbm{H}_0 =
      \left. \frac{\partial^2 F}{\partial\bfs{\sigma} \partial\bfs{\sigma}}
      \right|_{\bfs{\sigma}_0} , \quad
    n_r = \bfm{e}_r \cdot \bfm{n}_0 .
  \end{equation}
With this notation, the first derivatives of $\varrho_0$ and $F^\ast$,
Eqs.~(\ref{eq:drho0}) and (\ref{eq:dFs}), respectively, are written as
  \begin{equation}
    \frac{\partial \varrho_0}{\partial \bfs{\sigma}} =
      \frac{\varrho_0}{\varrho} \big( \bfm{e}_r - n_r^{-1} \bfm{n}_0 \big) , \quad
    \frac{\partial F^\ast}{\partial \bfs{\sigma}} =
      \frac{1}{\varrho_0} \big( n_r^{-1} \bfm{n}_0 \big) .
  \end{equation}

As discussed in Section~\ref{sec:general}, the implicit dependence of $\varrho_0$ on
$\bfs{\sigma}$ is introduced by the condition that $\bfs{\sigma_0}$ lies on the
yield surface, i.e., $F(\bfm{o}+\varrho_0(\bfs{\sigma})\bfm{e}_r(\bfs{\sigma}))=0$.
This equation is differentiated with respect to $\bfs{\sigma}$ twice, and the
resulting equation is solved for the unknown second derivative of $\varrho_0$, just
like in the case of the first derivative in Section~\ref{sec:general}, and the
following formula is obtained:
  \begin{eqnarray}
    \frac{\partial^2 \varrho_0}{\partial\bfs{\sigma} \partial\bfs{\sigma}}
      &=&
      \frac{\varrho_0}{\varrho^2} \Big(
      \mathbbm{S} - \bfm{e}_r \otimes \bfm{e}_r
      - 2 n_r^{-1} \big( \bfm{e}_r \otimes \bfm{n}_0 \big)_{\it sym}
      + 2 n_r^{-2} \bfm{n}_0 \otimes \bfm{n}_0
      \Big)
      \nonumber
      \\
      \label{eq:d2rho0}
      &&
      - \frac{\varrho_0^2}{\varrho^2} \Big(
      n_r^{-1} \mathbbm{H}_0
      - 2 n_r^{-2} \big( \bfm{n}_0 \otimes \mathbbm{H}_0 [ \bfm{e}_r ] \big)_{\it sym}
      + n_r^{-3} \big( \bfm{e}_r \cdot \mathbbm{H}_0 [ \bfm{e}_r ] \big)
        \bfm{n}_0 \otimes \bfm{n}_0
      \Big) .
  \end{eqnarray}
With this at hand, the second derivative of $F^\ast$ is finally obtained in a
remarkably simple form,
  \begin{equation}
    \frac{\partial^2 F^\ast}{\partial\bfs{\sigma} \partial\bfs{\sigma}}
      =
      \frac{1}{\varrho} \Big(
      n_r^{-1} \mathbbm{H}_0
      - 2 n_r^{-2} \big( \bfm{n}_0 \otimes \mathbbm{H}_0 [ \bfm{e}_r ] \big)_{\it sym}
       + n_r^{-3} \big( \bfm{e}_r \cdot \mathbbm{H}_0 [ \bfm{e}_r ] \big)
        \bfm{n}_0 \otimes \bfm{n}_0
      \Big) .
  \end{equation}


\bibliographystyle{spbasic}

\bibliography
{%
../../BIBTEX/contact,%
../../BIBTEX/lubrication,%
../../BIBTEX/stupkiewicz,%
../../BIBTEX/micromechanics,%
../../BIBTEX/phasetrans}

\begin{thebibliography}{19}

\providecommand{\natexlab}[1]{#1}

\providecommand{\url}[1]{{#1}}

\providecommand{\urlprefix}{URL }

\expandafter\ifx\csname urlstyle\endcsname\relax

  \providecommand{\doi}[1]{DOI~\discretionary{}{}{}#1}\else

  \providecommand{\doi}{DOI~\discretionary{}{}{}\begingroup

  \urlstyle{rm}\Url}\fi

\providecommand{\eprint}[2][]{\url{#2}}



\bibitem[{Bigoni(2012)}]{Bigoni12}
Bigoni D (2012) Nonlinear Solid Mechanics: Bifurcation Theory and Material
  Instability. Cambridge University Press, New York



\bibitem[{Bigoni and Piccolroaz(2004)}]{BigoniPiccolroaz04}
Bigoni D, Piccolroaz A (2004) Yield criteria for quasibrittle and frictional
  materials. Int J Sol Struct 41:2855--2878



\bibitem[{Bonnesen and Fenchel(1987)}]{BonnesenFenchel87}
Bonnesen T, Fenchel W (1987) Theory of convex bodies. BCS Associates, Moscow,
  Idaho, USA



\bibitem[{Brannon and Leelavanichkul(2010)}]{BrannonLeelavanichkul10}
Brannon RM, Leelavanichkul S (2010) A multi-stage return algorithm for solving
  the classical damage component of constitutive models for rocks, ceramics,
  and other rock-like media. Int J Fract 163:133--149



\bibitem[{Foster et~al(2005)Foster, Regueiro, Fossum, and Borja}]{Foster05}
Foster CD, Regueiro RA, Fossum A, Borja RI (2005) Implicit numerical
  integration of a three-invariant, isotropic/kinematic hardening cap
  plasticity model for geomaterials. Comp Meth Appl Mech Engng 194:5109--5138



\bibitem[{Hiriart-Urruty and Lemarechal(1993)}]{HiriartUrrutyLemarechal93}
Hiriart-Urruty JB, Lemarechal C (1993) Convex Analysis and Minimization
  Algorithms I: Fundamentals. Springer, Berlin Heidelberg



\bibitem[{Jeremic et~al(1999)Jeremic, Runesson, and Sture}]{Jeremic99}
Jeremic B, Runesson K, Sture S (1999) A model for elastic-plastic pressure
  sensitive materials subjected to large deformations. Int J Sol Struct
  36:4901--4918



\bibitem[{Korelc(2002)}]{Korelc02}
Korelc J (2002) Multi-language and multi-environment generation of nonlinear
  finite element codes. Engineering with Computers 18:312--327



\bibitem[{Korelc(2009)}]{Korelc09}
Korelc J (2009) Automation of primal and sensitivity analysis of transient
  coupled problems. Comp Mech 44:631--649



\bibitem[{Korelc and Stupkiewicz(2014)}]{KorelcStupkiewicz14}
Korelc J, Stupkiewicz S (2014) Closed-form matrix exponential and its
  application in finite-strain plasticity. Int J Num Meth Engng
  Doi:10.1002/nme.4653



\bibitem[{Nocedal and Wright(2006)}]{NocedalWright06}
Nocedal J, Wright SJ (2006) Numerical Optimization, 2nd edn. Springer, New York



\bibitem[{Penasa et~al(2014)Penasa, Piccolroaz, Argani, and Bigoni}]{Penasa14}
Penasa M, Piccolroaz A, Argani L, Bigoni D (2014) Integration algorithms of
  elastoplasticity for ceramic powder compaction. J Eur Cer Soc
  Doi:10.1016/j.jeurceramsoc.2014.01.041



\bibitem[{Perez-Foguet and Armero(2002)}]{PerezFoguetArmero02}
Perez-Foguet A, Armero F (2002) On the formulation of the closest-point
  projection algorithms in elastoplasticity---part {II}: {G}lobally convergent
  schemes. Int J Num Meth Engng 53:331--374



\bibitem[{Piccolroaz et~al(2006{\natexlab{a}})Piccolroaz, Bigoni, and
  Gajo}]{Piccolroaz06-1}
Piccolroaz A, Bigoni D, Gajo A (2006{\natexlab{a}}) An elastoplastic framework
  for granular materials becoming cohesive through mechanical densification.
  {P}art {I} -- small strain formulation. Eur J Mech A/Solids 25:334--357



\bibitem[{Piccolroaz et~al(2006{\natexlab{b}})Piccolroaz, Bigoni, and
  Gajo}]{Piccolroaz06-2}
Piccolroaz A, Bigoni D, Gajo A (2006{\natexlab{b}}) An elastoplastic framework
  for granular materials becoming cohesive through mechanical densification.
  {P}art {II} -- the formulation of elastoplastic coupling at large strain. Eur
  J Mech A/Solids 25:358--369



\bibitem[{Sheng et~al(2011)Sheng, Augarde, and Abbo}]{ShengAugardeAbbo11}
Sheng D, Augarde CE, Abbo AJ (2011) A fast algorithm for finding the first
  intersection with a non-convex yield surface. Comp Geotech 38:465--471



\bibitem[{Simo and Hughes(1998)}]{SimoHughes98}
Simo JC, Hughes TJR (1998) Computational Inelasticity. Springer-Verlag, New
  York



\bibitem[{de~Souza~Neto et~al(2008)de~Souza~Neto, Peric, and
  Owen}]{SouzaNeto08}
de~Souza~Neto EA, Peric D, Owen DRJ (2008) Computational Methods for
  Plasticity: Theory and Applications. Wiley



\bibitem[{Stupkiewicz et~al(2014)Stupkiewicz, Piccolroaz, and
  Bigoni}]{StupPicBig14}
Stupkiewicz S, Piccolroaz A, Bigoni D (2014) Elastoplastic coupling to model
  cold ceramic powder compaction. J Eur Cer Soc
  Doi:10.1016/j.jeurceramsoc.2013.11.017



\end{thebibliography}


\end{document}